\documentclass[twocolumn,amsmath,amssymb,showpacs,pra,superscriptaddress,nofootinbib,aps]{revtex4-1}

\usepackage{color}
\usepackage{array}
\usepackage{graphicx}
\usepackage{bm}

\usepackage{hyperref}
\usepackage[english]{babel}
\usepackage[utf8]{inputenc}
\usepackage[stable]{footmisc}
\usepackage{booktabs}
\usepackage{multirow}
\usepackage{longtable}
\usepackage{afterpage}
\usepackage{amsmath}
\usepackage{float}
\usepackage{physics}
\usepackage{amsfonts}

\usepackage{amssymb}
\usepackage{braket}
\usepackage{graphicx}
\usepackage{ulem}
\usepackage{todonotes}
\usepackage{subfigure}
\usepackage{bm}
\usepackage{xcolor}
\usepackage{lipsum}

\begin{document}
\setlength{\abovedisplayskip}{6pt}
\setlength{\belowdisplayskip}{6pt}

\title{In-medium bound states of two bosonic impurities in a one-dimensional Fermi gas}
\author{D. Huber}
\email{dhuber@theorie.ikp.physik.tu-darmstadt.de}
\affiliation{Institut f\"ur Kernphysik, Technische Universit\"at Darmstadt,
64289\ Darmstadt, Germany}
\author{H.-W. Hammer}
\email{hammer@theorie.ikp.physik.tu-darmstadt.de}
\affiliation{Institut f\"ur Kernphysik, Technische Universit\"at Darmstadt,
64289\ Darmstadt, Germany}
\affiliation{ExtreMe Matter Institute EMMI, GSI Helmholtzzentrum f\"ur 
Schwerionenforschung, 64291\ Darmstadt, Germany}
\author{A. G. Volosniev}
\email{artem.volosniev@ist.ac.at}
\affiliation{Institut f\"ur Kernphysik, Technische Universit\"at Darmstadt,
64289\ Darmstadt, Germany}
\affiliation{Institute of Science and Technology Austria, Am Campus 1, 3400 Klosterneuburg, Austria}
\date{\today}

\begin{abstract}
  We investigate the ground-state energy of a one-dimensional Fermi gas with two bosonic impurities. We consider spinless fermions with no 
fermion-fermion interactions. The fermion-impurity and impurity-impurity interactions are modelled with Dirac delta functions.
  First, we study the case where impurity and fermions have equal masses, and the impurity-impurity two-body interaction is identical to the fermion-impurity interaction, such that
  the system is solvable with the Bethe ansatz. For attractive interactions, we find that the energy of the impurity-impurity
  subsystem is below the energy of the bound state that exists without the Fermi gas.  We interpret this as a manifestation of attractive boson-boson interactions induced by the fermionic medium, and refer to the impurity-impurity subsystem as an in-medium bound state. For repulsive interactions, we find no in-medium bound states. Second, we construct an effective model to describe these
  interactions, and compare its predictions to the exact solution. We use this effective model
  to study non-integrable systems with unequal masses and/or potentials. We discuss parameter regimes for which impurity-impurity attraction induced by the Fermi gas can lead to the formation of in-medium bound states
made of bosons that repel each other in the absence of the Fermi gas.
\end{abstract}

\smallskip
\maketitle

\section{Introduction}
An environment with mobile impurity atoms is a cherished model system in quantum 
physics. It is a testbed for introducing and testing quasiparticle concepts, e.g.,
polarons and bipolarons, which naturally appear when studying 
the movement of electrons in crystals~\cite{landau1948,pekar1951, mott1995}, 
$^3$He atoms in superfluid $^4$He~\cite{baym2008}, or even protons in neutron matter~\cite{kutschera1993}. 
Nowadays, these concepts can be examined using quantum simulators based upon ultracold atoms~\cite{zwierlein2009,salomon2009, massignan2014, hu2016,arlt2016, schmidt2018}. An important topic that can be addressed with cold atom systems
is the physics of impurity-impurity correlations induced by a medium~\cite{bruderer2007,schecter2014,Keiler_2018,
 naidon2018,volosniev2018,bruun2018a,bruun2018,pavlov2018, mistakidis2019,mistakidis2019_two,petkovic2019, chin2019}.
This topic is relevant for basic research, and in applications motivated by bound states of dressed 
electrons and their relation to high-T\textsubscript{c} superconductors~\cite{mott1994},\footnote{
Experiments with ultracold atoms can give important insight  
into the physics of induced attractive potentials between dressed electrons. However, atoms interact via short-range potentials limiting the relationship between impurities in cold atoms and dressed electrons.
For example, the predicted abrupt change
of the mean distance between two polarons across the unbound-polarons to bipolaron
transition~\cite{adamovski1989} most probably cannot be simulated in these experiments because the Coulomb interaction is an important ingredient for observing this effect~\cite{schmickler2019}.}.

In this paper, we calculate the ground-state energy of a one-dimensional (1D) Fermi gas with two bosonic impurities, see Fig.~\ref{visual}. 
One-dimensional geometries typically enhance interaction effects~\cite{takada1982} opening up
the possibility of observing bound states supported by the induced attraction~\footnote{Note that any attractive potential supports at least one bound state in the 1D world~\cite{landaubook}}.  
Another feature that separates one spatial dimension from higher dimensions is 
the long-range tail of correlations. For example, Friedel oscillations~\cite{Friedel1958} decay as $\sim 1/r^D$ where $D$ is the dimension of space.  These enhanced correlations may be useful to simulate many-body phenomena beyond short-range physics typical for cold atoms.

Our paper is organized as follows. We start by 
introducing the Hamiltonian of our 1D model in Section~\ref{sec:formulation}. In
Section~\ref{sec:exactResults}, we study impurity-impurity correlations in
the limiting case of equal masses, $M=m$. All interactions 
are identical and parametrized by Dirac delta functions. The fermions do not interact among each each other due to the Pauli exclusion principle.
The system is solvable by the Bethe ansatz,
which is a common starting point for analyzing cold atom systems in 1D geometries~\cite{rigol2011, guan2013, Batchelor_2016}.
In Section~\ref{sec:eff_models}, we go on to discuss effective models for describing two impurities 
in a medium and benchmark them against the Bethe ansatz results.
Afterwards, we use the effective models to investigate non-integrable systems; our focus is on the appearance of in-medium bound states. 
We discuss the transition from unbound impurities to bound impurities, which can be tested in cold atom experiments, for example, by measuring the spectroscopic response
or by studying the collapse dynamics~\cite{chin2019} in imbalanced Bose-Fermi mixtures.
Finally, Section~\ref{sec:summ} contains a summary
of our results and an outlook.

\section{Formulation}
\label{sec:formulation}
\begin{figure}[t]
\centering
\includegraphics[scale=0.8]{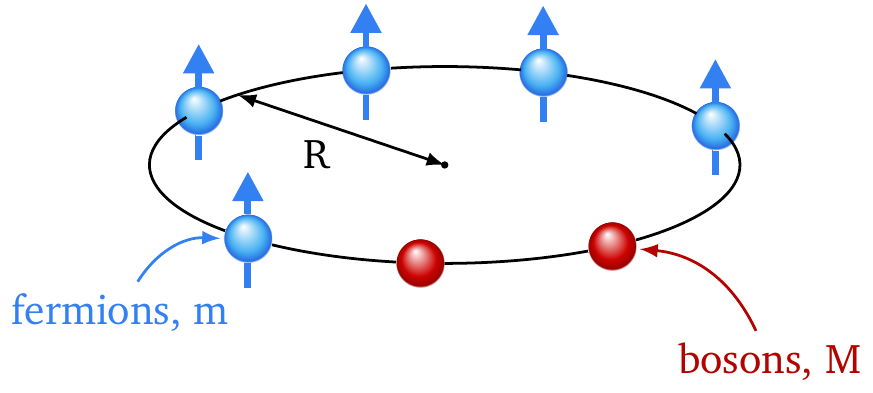} 
\caption{An illustration of the system:
Two bosonic impurities in a one-dimensional Fermi gas. 
Periodic boundary conditions are employed, i.e., the system 
lives on a ring of radius $R$. The mass of an impurity (fermion) is
denoted by $M$ ($m$).}
\label{visual}
\end{figure}
We consider two bosonic impurities interacting 
via a zero-range potential with $N_f$ spinless (fully spin-polarized) fermions. 
For convenience, we assume that $N_f$ is an odd number. 
This assumption does not limit the generality of our results
as we are interested in the limit $N_f\to \infty$.
 Note that the bosons may, in principle, possess spin quantum numbers. However, the bosonic spin part of the wave function is not important 
for our discussion, since we focus on the ground state. 
It is only necessary that the wave function is symmetric with respect to the exchange of the spatial coordinates of the impurity atoms.
The particles are confined to a ring of radius $R$, see Fig.~\ref{visual}.
The Hamiltonian for the system reads
\begin{equation}
H=H_f+H_b+V_{fb},
\label{eq:ham}
\end{equation}
where $H_f$ describes fermions:
\begin{equation}
H_f=-\sum_{j=1}^{N_f} \frac{\hbar^2}{2m}\frac{\partial^2}{\partial X_j^2},
\end{equation}
with $m$ the fermion mass.
$H_b$ describes the impurity bosons:
\begin{equation}
H_b=-\frac{\hbar^2}{2M}\frac{\partial^2}{\partial Y_1^2}-\frac{\hbar^2}{2M}\frac{\partial^2}{\partial Y_2^2}+g_{II}\delta(Y_1-Y_2),
\label{eq:h_b}
\end{equation}
where $M$ is the boson mass, and $g_{II}$ is the strength of the boson-boson interaction. 
The interaction between fermions and bosons is written as
\begin{equation}
V_{fb}=g\sum_{i,j}\delta(Y_i-X_j), \quad i=1,2;\: j=1,\ldots,N_f,
\end{equation}
where $g$ is the corresponding interaction strength.
We solve the Schr{\"o}dinger equation $H\psi=\epsilon \psi$ for the ground state for different $N_f$ and sizes $L=2\pi R$ of the system.
Then, we extrapolate the energies to the thermodynamic limit,
$N_f,L \rightarrow \infty$, assuming a fixed density of the Fermi gas, $N_f/L = \rho$.
For convenience, we introduce the dimensionless quantities $y_j=Y_j\rho$, 
$x_j = X_j \rho$, $l=L\rho$, $c_{II}=m g_{II}/(\hbar^2 \rho)$, $c=m g/(\hbar^2 \rho)$, and
$\varepsilon= 2m \epsilon/(\hbar^2 \rho^2)$.

 Note that working with a finite number of particles allows us also to study 
the transition of the energy from a few- to many-body limits, an exciting research venue,
which can be addressed with cold atoms~\cite{wenz2013, zinner2016,Rammelmuller2017a}. 
We observe that the ground state energies of systems with $N_f$ of the order of 10-20 particles can be accurately described 
using the results derived in the thermodynamic limit.

\section{Solvable Limits}
\label{sec:exactResults}

\subsection{Bethe-Ansatz-Solvable Case}

First, we consider the most symmetric case: $c_{II}=c$ and $m=M$,
whose Hamiltonian we write as
\begin{equation}
h_{BA}=-\sum_{j=1}^{N}\frac{\partial^2}{\partial x_j^2}+2c\sum_{j<l}^{N}\delta \left( x_j-x_l \right)\,,
\label{eq:hba}
\end{equation}
where $N=N_f+2$; we set $x_{N - 1}=y_1$ and $x_N=y_2$ to explicitly demonstrate the particle exchange symmetry. 
The ground state of $h_{BA}$ with fermions at the coordinates ($x_1,...,x_{N_f}$) and bosons at ($x_{N-1}, x_N$)
can be studied experimentally with SU(3)-symmetric fermions, e.g., with $^{173}$Yb~\cite{fallani2014}.
Indeed, the ground state of SU(3) fermions with the particle decomposition $N_f+1+1$ has a bosonic symmetry for the exchange of the particles in the $1+1$ subsystem.  
To understand this, note that: (i) the particles in the $1+1$ subsystem are distinguishable particles and, hence, there exist no apriori symmetry requirements for their exchange; (ii) the Hamiltonian $h_{BA}$ commutes with the particle exchange operator; (iii) the bosonic symmetry leads to the lowest energy.
Furthermore, we expect that the ground state energies of isotopic systems with a small mass imbalance, e.g., $^6$Li-$^7$Li and $^{39}$K-$^{40}$K (cf.~\cite{Ferrier-Barbut1035,wu2011}), can  be accurately described by Eq.~(\ref{eq:hba}).

The spectrum of the Hamiltonian $h_{BA}$ can be found
using the Bethe ansatz (BA)~\cite{yang1967}. Let us briefly summarize this approach.
For every ordering of particles (e.g., for $x_1<x_2<...<x_N$),
the wave function is written as a sum of the plane waves $e^{i\sum_j k_j x_j}$. 
For this wave function to fulfill the boundary conditions at $x_i=0$, $x_i=l$ and $x_i=x_j$ for all $i$ and $j$,
the quasi-momenta $k_j$ must satisfy the BA equations 
\begin{align}
\begin{split}
& e^{ik_jl}=\frac{k_j-\Lambda_1+\frac{ic}{2}}{k_j-\Lambda_1-\frac{ic}{2}} 
\frac{k_j-\Lambda_2+\frac{ic}{2}}{k_j-\Lambda_2-\frac{ic}{2}}\,,\quad 1\leq j\leq N;\\
& \prod_{j=1}^N\frac{k_j-\Lambda_1+\frac{ic}{2}}{k_j-\Lambda_1-\frac{ic}{2}}=1\,,\quad 
\prod_{j=1}^N\frac{k_j-\Lambda_2+\frac{ic}{2}}{k_j-\Lambda_2-\frac{ic}{2}}=1\,;
\end{split}
\label{eq:BA}
\end{align}
where the bosonic and fermionic symmetries have already been implemented~\cite{yang1971,guan2005,demler2006}. 
$\Lambda_1$ and $\Lambda_2$ are to be determined together with the set $\{k_j\}$. 
Once the BA equations are solved, the energy of the system is 
determined as $\varepsilon=\sum_{j=1}^N k_j^2$.
Note that the number of unknowns in Eqs.~(\ref{eq:BA}) for the ground state
can be reduced to $(N_f+3)/2$ from $N_f+4$.  
Indeed, the total (angular) momentum must be zero in the ground state,
$\sum_j k_j=0$. This together with the fact that the wave function is real makes
the  quasi-momenta appear in pairs (we exemplify this below). In addition, one can show that $\Lambda_1 =-\Lambda_2$. 

To solve Eq.~(\ref{eq:BA}) for the ground state, we apply Newton's method, which requires 
an accurate initial estimate of $k_j$ and $\Lambda_j$. For $c\to 0$, we obtain
this estimate directly from the BA equations (see~Appendix~\ref{app:a}):
\begin{align}
\begin{split}
k_1 & \simeq \sqrt{\frac{3c}{l}}\,,\quad  k_2\simeq -\sqrt{\frac{3c}{l}} \,,\quad k_3\simeq 0\,,\\
k_j & \simeq k_j^{(0)}+\frac{2 c}{k_j^{(0)} l}\,,\quad\text{for }4\leq j\leq N \,;\\
\Lambda_1 & \simeq \sqrt{\frac{c}{l}}\,,\quad \Lambda_2\simeq-\sqrt{\frac{c}{l}},
\end{split}
\label{eq:initial_weak}
\end{align}
where $k_j^{(0)}$ is the quasi-momenta at $c=0$.
Note that in Eq.~(\ref{eq:initial_weak}) the quasi-momenta are related pairwise, e.g., $k_1=-k_2$, as has already been mentioned. 
The only non-paired quasi-momentum is $k_3=0$. 
Estimate~(\ref{eq:initial_weak}) allows us to calculate $\{k_i\}$
and obtain solutions for weak interactions, $|c|\ll1$.
We then follow the solutions as $c$ is changed in small steps.
An initial guess for moderate interactions is obtained from a Taylor series constructed 
using solutions at smaller values of $|c|$, see Appendix~\ref{app:a0}.

We solve Eq.~(\ref{eq:BA}) for a sequence of $N_f$ and extrapolate to the thermodynamic limit. To this end, we subtract from the energy the zero-interaction offset and fit the difference with $\varepsilon (c)-\varepsilon (0) = \varepsilon_{\infty}+A_1/N+A_2/N^2$,
where $\varepsilon_{\infty}, A_1$ and $A_2$ are fitting parameters.
It is straightforward to argue for the form of the fitting function, $\varepsilon_{\infty}+A_1/N+A_2/N^2$,
in the case of strong interactions ($c\to \pm \infty$) for which the energies can be calculated 
using a non-interacting Fermi gas. 
We do not attempt to validate the fitting function for finite values of $c$, since
we observe that the form of the function is not important for our analysis~(see Appendix~\ref{app:b}).

\begin{figure}
\centering
\includegraphics[width=0.9\linewidth]{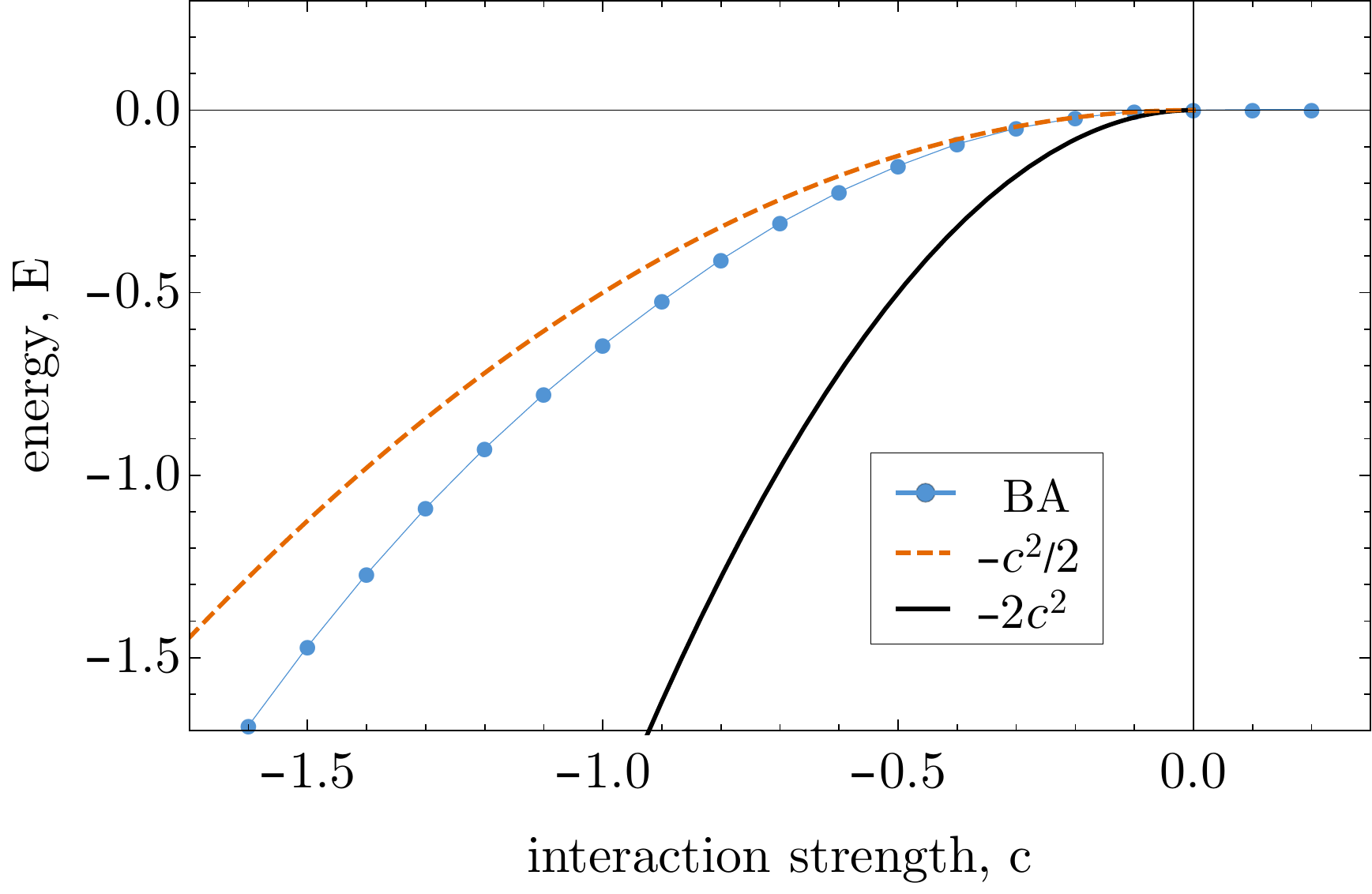} 
\caption{The (blue) dots show the in-medium binding energy of two bosonic impurities in a Fermi gas for the Bethe-ansatz-integrable case,
the size of the dots can be used as an estimate for an error from the fitting procedure used
to obtain the thermodynamic limit (see~Appendix~\ref{app:b}).
The solid (blue) curve is added to guide the eye.
The dashed (orange) curve shows the binding energy of two bosons without the Fermi gas.
We interpret the gap between the dashed curve and the dots as a manifestation of
attractive impurity-impurity interactions mediated by the Fermi gas.
For comparison, we also present the ground state energy of two bosons and one fermion~\cite{mcguire1964}, 
see the solid (black) curve.}
\label{fig:Bindenergy}
\end{figure}

To investigate induced correlations in the thermodynamic limit, we introduce the ``in-medium binding energy"
$E=\varepsilon_{\infty}-2\mathcal{E}$, where $\mathcal{E}$ is the energy gain for immersing
one impurity in a Fermi gas~\cite{McGuire1965,McGuire1966}~(see Appendix~\ref{app:c}). The quantity $2\mathcal{E}$ describes the energy of two non-correlated impurities.
$E$ is presented in Fig.~\ref{fig:Bindenergy}. Note that $E$ cannot be positive, since any induced correlations between impurities must vanish when they are far apart. 
If $E=0$ the two impurities are not correlated, in general, they are infinitely far from each other. In this case, we say that there is no in-medium bound state, whereas if $E<0$ then there is at least one. 
Next, we analyze cases with $c<0$ and $c>0$ separately. 

{\it Repulsive interactions, $c>0$}. We calculate the energy for $c\lesssim 2$ and find that $E=0$ (within numerical accuracy), 
which means that there are no in-medium bound states. 
In principle, an in-medium bound state is possible for two impurities
  that repel each other without the Fermi gas. 
  This can happen if the energy cost of
  embedding an impurity in the medium is larger than the cost of bringing
  the impurities close together. Our calculations show that an in-medium bound state does not appear when $c_{II}=c>0$. For $c\to 0$ it can easily be understood that $E=0$,
  since the impurity-impurity interaction without the Fermi gas scales
as $c$ (see~Eqs.~(\ref{eq:h_b}) and (\ref{eq:hba})), whereas the induced impurity-impurity interaction is expected to scale as $c^2$ (see Sec.~\ref{sec:eff_models}). 
Therefore, the interaction volume, i.e., the space integral of the effective impurity-impurity interaction, is necessarily a small positive quantity for $c\to 0$, which does not allow for the existence of a
bound state~\cite{landaubook, simon1976}. 

In the limit of strong interactions some extra insight can also be gained. 
For $c\to \infty$ the important degrees of freedom are spins~\cite{volosniev2014_nat,santos2014,volosniev2015,massignan2015},
which allows one to map the Hamiltonian $h_{BA}$ onto an XX spin chain~\cite{santos2017}
with constant coefficients,
\begin{equation}
h_{BA}\to -\frac{J}{2}\sum(\sigma_x^i\sigma_x^{i+1}+\sigma_y^i\sigma_y^{i+1}),
\end{equation}
where $\sigma_{x}^i$ and $\sigma_y^i$ are the Pauli matrices acting on the spin 
at site $i$; $J$ is an exchange coefficient proportional to $1/c$, see \cite{volosniev2017} for the derivation in a homogeneous environment. 
The system in Fig.~\ref{visual} with $c\to \infty$ is then identical to a linear spin chain with two spin deviates (magnons) 
for which a bound state is not expected~\cite{Hodgson_1984}.

{\it Attractive interactions, $c<0$}. Figure~\ref{fig:Bindenergy} shows that for $c<0$ there is an in-medium bound state 
whose energy is below the ground state energy of the Hamiltonian that describes two bosons without the Fermi gas, i.e., $H_b$ from Eq.~(\ref{eq:h_b}). 
This lowering of the energy is a manifestation of the induced impurity-impurity correlations, which 
we interpret in Sec.~\ref{sec:eff_models} using an effective impurity-impurity potential.
Since we are interested in interactions mediated by the Fermi gas,
we do not discuss in this paper the limit $c\to -\infty$, which implies the formation of a tightly bound trimer (in analogy to trions in SU(3) symmetric Fermi gases~\cite{guan2013}).
This trimer is supported by the fundamental (not induced) interaction, and, hence, is beyond the scope of this paper. 
For the sake of discussion, we present the energy of a trimer in Fig.~\ref{fig:Bindenergy}. We expect that this energy, $-2 c^2$, determines 
the behavior of the system as $c\to-\infty$. 
In the rest of the paper, we only explore $c\gtrsim-2$.

\subsection{Two Static Impurities}
\label{sec:static_imp}
\begin{figure}[t]
\centering
\includegraphics[width=0.9\linewidth]{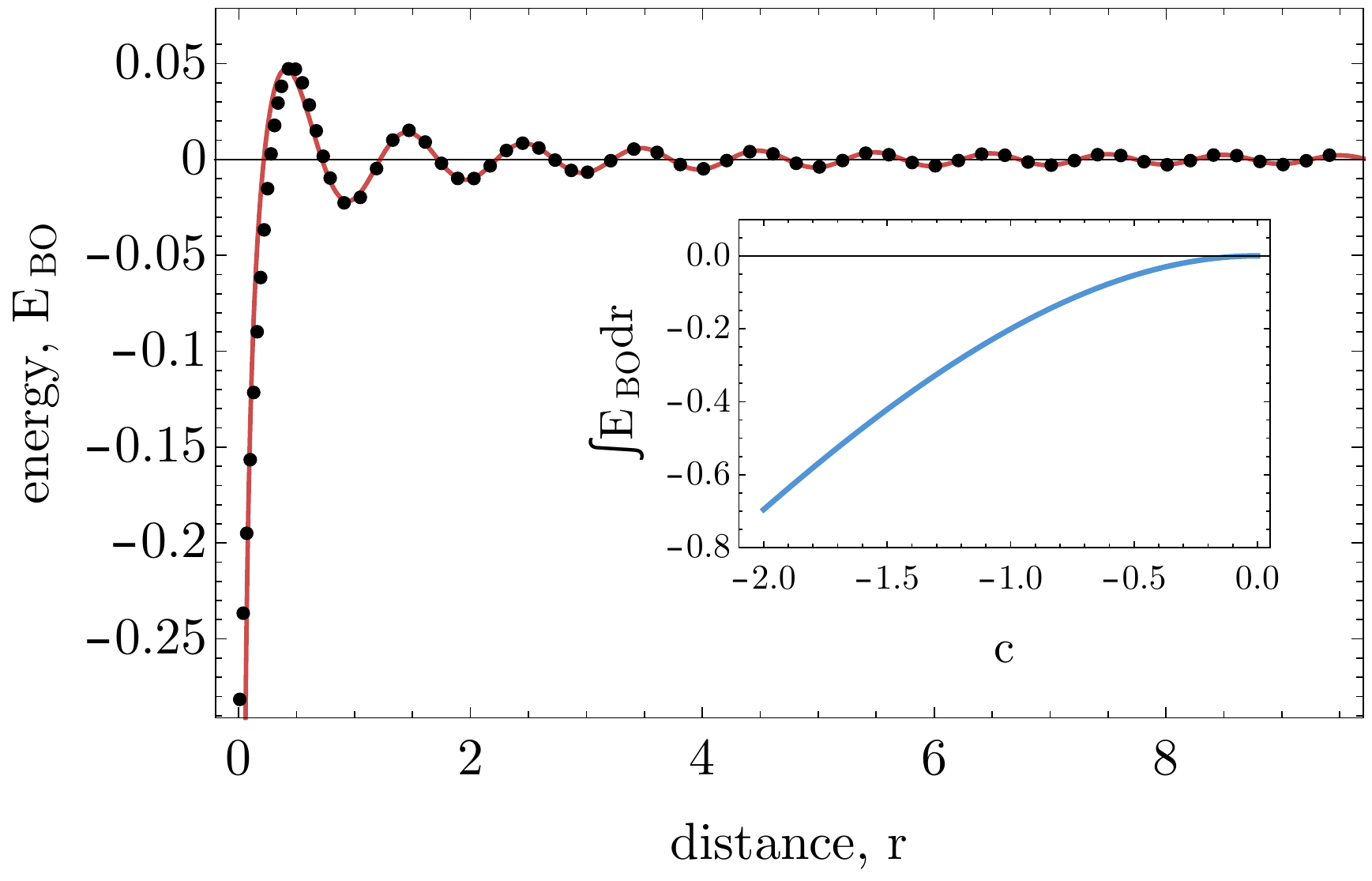} 
\caption{The energy of a Fermi gas with two static impurities. The dots show our results for $c=-0.5$.
The curve shows the fit to the long-range tail~\cite{zwerger2005,zwerger2007} $B\cos(A r+\delta)/r$ for $c\to 0$, where $A,B$ and $\delta$ are fitting parameters.  Note that the fit is accurate almost everywhere, except the region of $r\to 0$, where the exact result must be used. The inset present the integral of $E_{BO}$ over the entire space.}
\label{fig:potentialAndIntegralPlot}
\end{figure}
Before we discuss effective models that describe two mobile impurities in a sea of fermions,
we consider two static impurities $M\to \infty$ -- another analytically solvable limit of the Hamiltonian~(\ref{eq:ham}).
The solution is obtained by solving the one-body problem:
one particle in a ring with a potential due to the two impurities fixed at $-r/2$ and $r/2$
\begin{align}
\label{eq:hamStaticImp}
h_{BO}=-\frac{\partial^2}{\partial x^2} + 2c\left[ \delta \left(x-\frac{r}{2} \right)+\delta \left(x+\frac{r}{2} \right)\right]\,,
\end{align}
where the subscript emphasizes the connection to the Born-Oppenheimer (BO) approximation,
which will be employed below. The spectrum of $h_{BO}$ depends on the distance $r$. We calculate this dependence only for attractive interactions, i.e., $c<0$; the repulsive case can be calculated in a similar manner. To obtain the ground state energy of the Fermi gas, $\varepsilon_{BO}(c,r)$,
we add the energies of the lowest $N_f$ eigenstates of the Hamiltonian $h_{BO}$~(see Appendix~\ref{app:d}).
The thermodynamic limit is calculated by extrapolating the results for systems with different values of $N_f$ and $l$ and a fixed ratio $N_f/l$. 
We observe that already for $N_f \simeq 19$ the energy for small values of $r$ can be used (for our discussion) as the energy in the thermodynamic limit. 
The solution for $r\to \infty$ can be obtained by fitting to the known form of the tail~\cite{zwerger2005,zwerger2007}. 

Figure~\ref{fig:potentialAndIntegralPlot} illustrates the energy $E_{BO}(r)=\varepsilon_{BO}(c,r)-\varepsilon_{BO}(0,r)-2 \mathcal{E}_{static}$ for $c=-0.5$
(we assume that $g_{II}=0$ for the sake of discussion).
$\mathcal{E}_{static}$ is the energy gain for immersing a single static impurity in a Fermi gas~\cite{parisi2017}
\begin{equation}
\mathcal{E}_{static}(c)=\left(\pi+\frac{c^2}{\pi}\right)\arctan(\frac{c}{\pi})+c-\frac{c^2}{2}.
\end{equation} 
The quantity $E_{BO}$ has a deep minimum at $r=0$ given by $\mathcal{E}_{static}(2c) - 2 \mathcal{E}_{static}(c)$
and an oscillatory tail. To derive this limiting value, note that 
when both impurities are at $r=0$, they act as a single impurity with the strength $2c$.  For $c\to 0$ the tail can be written simply as $c^2\cos(A r+\delta)/r$,
where $A$ and $\delta$ are constants. This tail can be obtained from Friedel oscillations~\cite{Friedel1958}.\footnote{An alternative derivation
  is given in Refs.~\cite{zwerger2005,zwerger2007}.} 
These oscillations determine the density of the Fermi gas at the position of the second impurity, provided that the first impurity is separated by the distance $r$. This density in turn determines the energy of the system,
according to first order perturbation theory.
 It is worthwhile noting that the dependence of $E_{BO}$
on $r$ is observable. It can, in principle, be probed in cold atom experiments by spectroscopy~\cite{zwerger2005}.

\section{Effective model}
\label{sec:eff_models}

\subsection{Bethe-ansatz-solvable system}
\label{sec:eff_BA}

\begin{figure}[htb]
\centering
\includegraphics[width=0.9\linewidth]{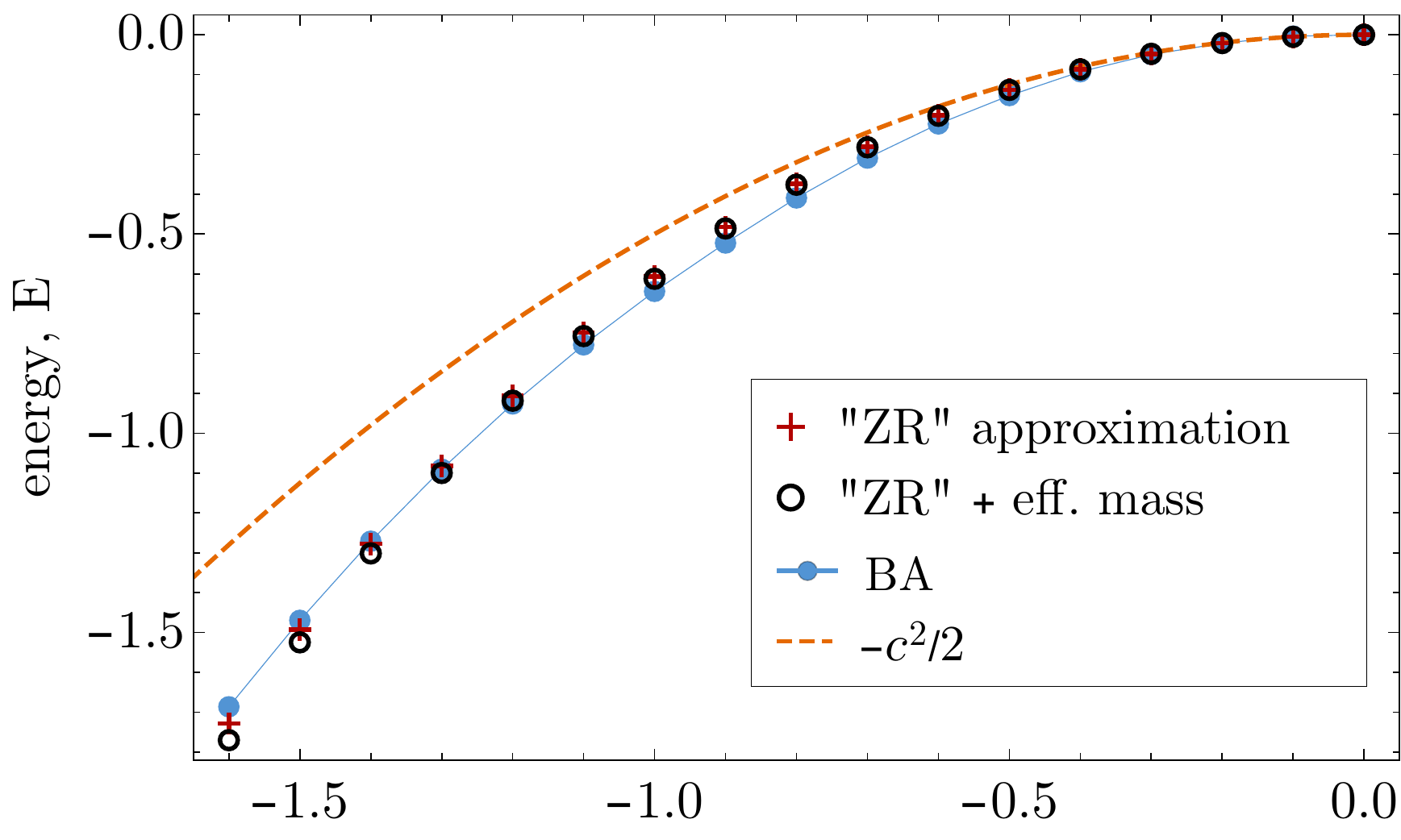} 
\includegraphics[width=0.9\linewidth]{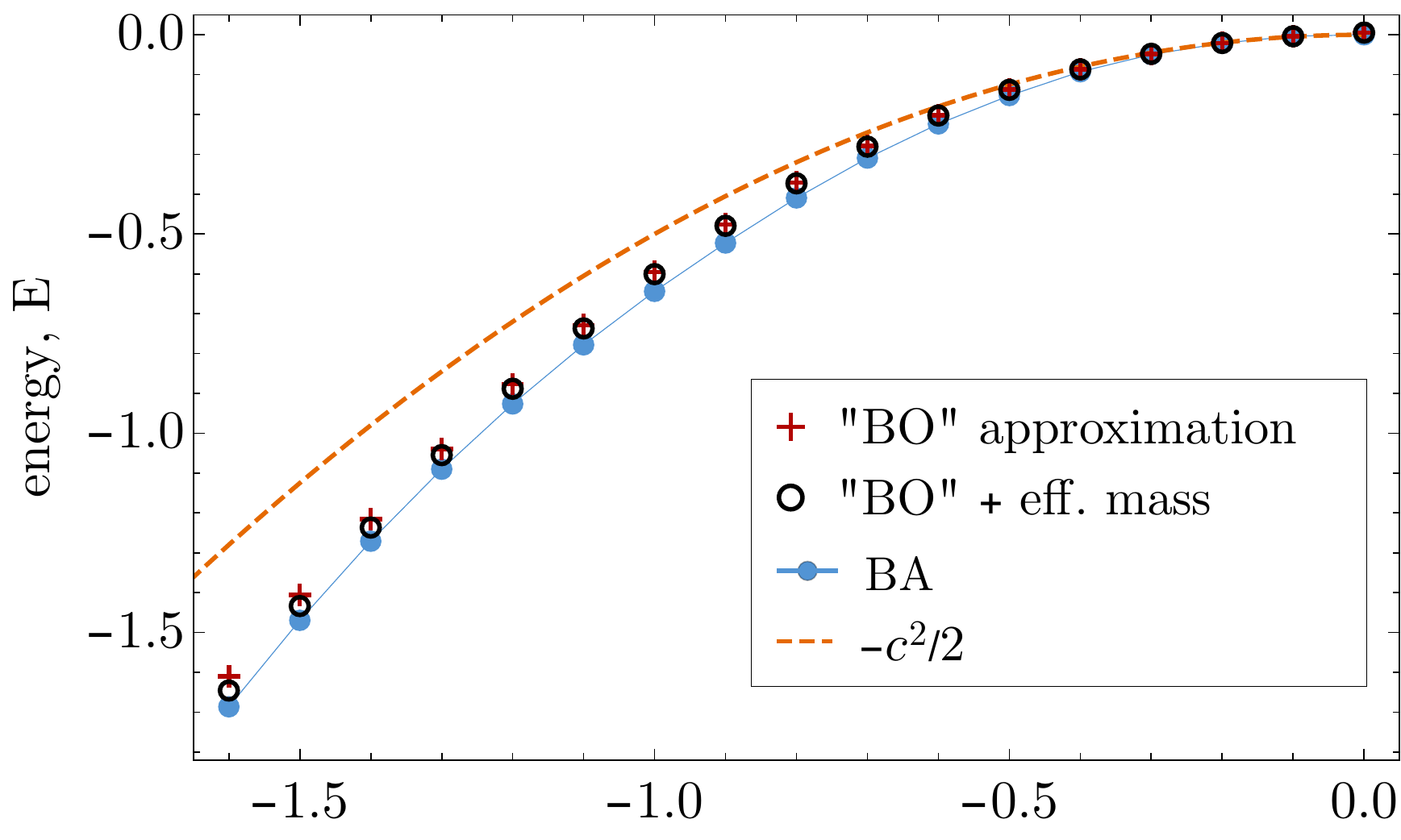} 
\includegraphics[width=0.9\linewidth]{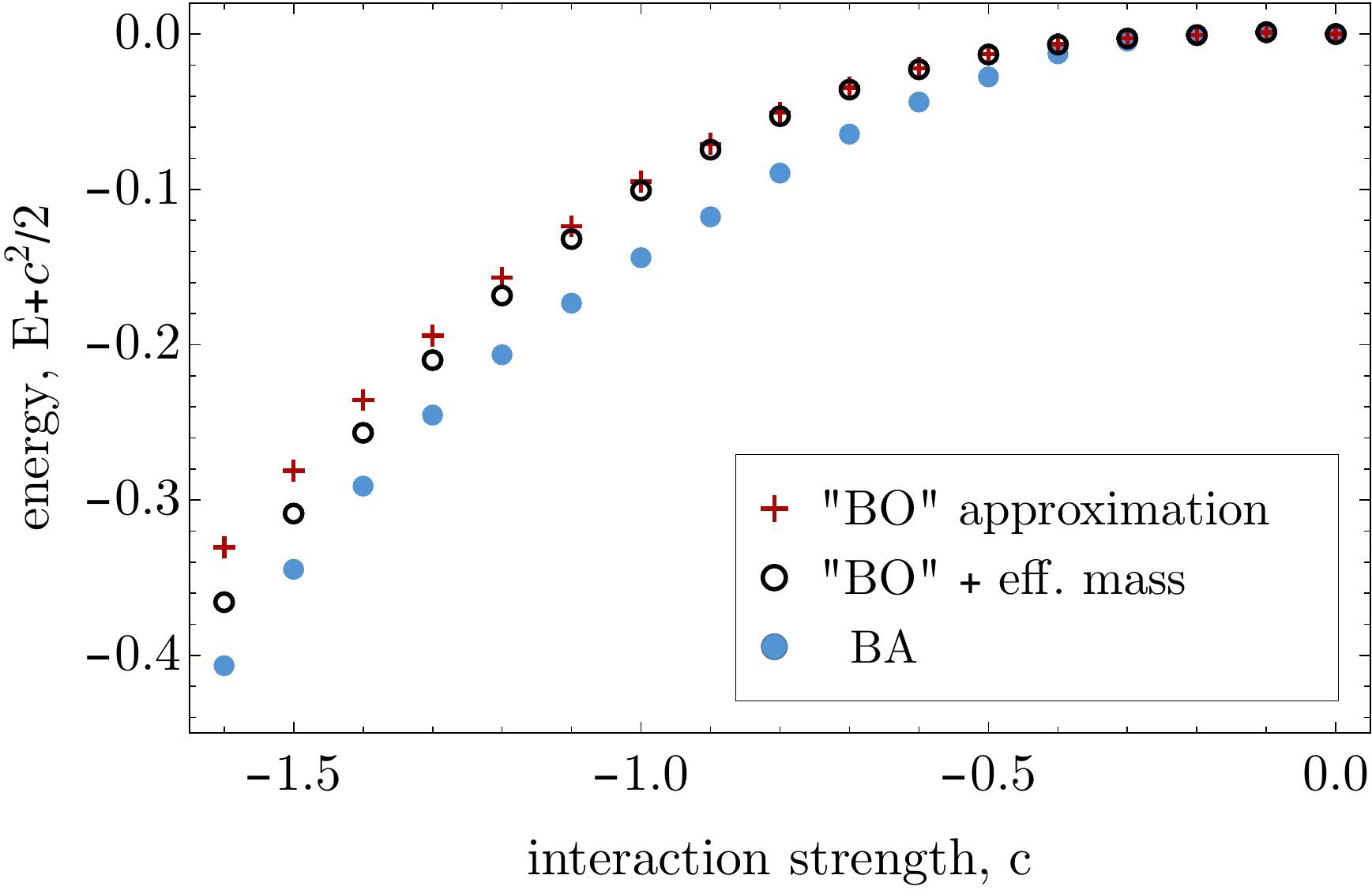} 
\caption{The in-medium binding energy of two impurities in a Fermi gas
compared to the zero-range and Born-Oppenheimer approximations. 
The dots show the exact BA results as in Fig.~\ref{fig:Bindenergy}, while
the dashed curve shows the binding energy of two bosons without the Fermi gas.
Top: Comparison to the zero-range approximation,
Eq.~(\ref{eq:Eeff_ZR}), with $m_{\mathrm{eff}}$ from 
Eq.~(\ref{eq:mass_renorm}) (circles) and
the result for $m_{\mathrm{eff}}=m$ (crosses).
Center: Comparison to the Born-Oppenheimer approximation with
the potential from Fig.~\ref{fig:potentialAndIntegralPlot} and
$m_{\mathrm{eff}}$ from Eq.~(\ref{eq:mass_renorm}) (circles) and
the result for $m_{\mathrm{eff}}=m$ (crosses).
Bottom: The data from the center panel shifted by $c^2/2$
and displayed on a larger scale. 
}
\label{fig:Finalresult}
\label{fig:MassImBal}
\end{figure}

 The ground state of a one-dimensional
Fermi gas ($N_f\to \infty$) with a single interacting impurity is orthogonal (for all non-zero values of interaction strengths) 
to the ground state of the corresponding non-interacting system.
This phenomenon is related to the Anderson orthogonality catastrophe~\cite{Anderson1967}. 
For the SU(2) symmetric case it can be conveniently studied using the BA equations~\cite{zotos1993}.
This orthogonality reduces the applicability of the polaron picture. 
For example, the dynamics after a sudden change of parameters cannot be captured 
by a single quasiparticle, instead it requires a continuum of states. 
Still, the notion of the effective mass and self-energy can be used to
describe the low-energy spectrum of a Fermi gas with one impurity~\cite{McGuire1965, McGuire1966, zotos1993, Combescot2009,guan2016},
suggesting the use of the following Hamiltonian to model the binding energy for the system of two impurities 
\begin{align}
h_{\mathrm{eff}}=-\frac{m}{m_{\mathrm{eff}}}\frac{\partial^2}{\partial y_1^2}-\frac{m}{m_{\mathrm{eff}}}\frac{\partial^2}{\partial y_2^2}+W(y_1-y_2)\,,
\label{eq:ham_eff}
\end{align}
where $m_{\mathrm{eff}}(c)$ is the effective mass of the impurity whose analytical form is presented in Refs.~\cite{McGuire1966,zotos1993,Combescot2009}, so that 
the Hamiltonian $h_{\mathrm{eff}}$ with $W=0$ correctly describes the low-energy spectrum of two non-interacting 
impurities. For simplicity, we work with the expansion of the effective mass truncated at the order $O(c^4)$,
\begin{equation}
\frac{m_{\mathrm{eff}}}{m}\simeq 1+\frac{c^2}{\pi^4}+\left(\frac{2}{\pi^6}-\frac{1}{6\pi^4}\right)c^3+\left(\frac{4}{\pi^8}-\frac{1}{2\pi^6}\right)c^4.
\label{eq:mass_renorm}
\end{equation} 
The terms of the order $O(c^5)$ can be neglected for interactions considered here, $c\in (-2,0)$.
This equation shows that the mass does not increase by more than a few
percent for the considered parameters.

The function $W$ in Eq.~(\ref{eq:ham_eff}) describes the impurity-impurity correlations. 
 We choose it as 
\begin{equation}
W(y_1-y_2)=2c\delta(y_1-y_2)+V(y_1-y_2),
\label{eq:effpot}
\end{equation}
where the first term is the interaction between impurities without the Fermi gas; the second term is an effective interaction mediated by the environment.
Note that the exact shape of $V$ is not required for our purpose.  We are interested in the weak and moderate interaction
regimes for which the knowledge of a few integrated properties of $V$ is sufficient, e.g., 
only $\int V(y)\mathrm{d}y$ determines the binding energy for $c\to 0$. Indeed, 
the ground state energy $E_{\mathrm{eff}}$ of $h_{\mathrm{eff}}$ for weak interactions reads~\cite{landaubook,simon1976}
\begin{equation}
E_{\mathrm{eff}}\simeq-\frac{m_{\mathrm{eff}}}{8m}\left[\int_{-\infty}^{\infty}W(y)\mathrm{d}y\right]^2,
\label{eq:Eeff_weak}
\end{equation}
     where the integration is over the whole system in the
      thermodynamic limit. This expression can be expanded as
\begin{equation}
E_{\mathrm{eff}}=-\frac{c^2}{2}-\frac{c}{2}\int V(y)\mathrm{d}y+ O(c^4),
\label{eq:expansion}
\end{equation}
where we assume that $\int V \mathrm{d}y$ scales as $c^2$ (see below) to estimate the neglected pieces.
Note that the renormalization of mass enters in $O(c^4)$ meaning that this effect 
may be disregarded for $c\to 0$ when compared to the effect of the two-body effective interaction.
The first term in Eq.~(\ref{eq:expansion}) is the ground state energy of two particles 
without the Fermi gas. The integral $\int V \mathrm{d}y$ must be negative to ensure that
the energy of the in-medium bound state is below $-c^2/2$ as in Fig.~\ref{fig:Bindenergy}.
Therefore, the effective interaction must be overall attractive. 
Let us discuss the two (arguably) simplest approximations that can be used to
appropriately choose $V$.

{\it Zero-range Potential}.
The most basic form of $V$ in Eq.~(\ref{eq:effpot}) consistent with the first two terms of 
the expansion~(\ref{eq:expansion}) is the zero-range (ZR) potential
\begin{equation}
V_{ZR}(y_1-y_2)=-\kappa \delta(y_1-y_2),
\label{eq:zero_range_pot}
\end{equation}
where $\kappa \equiv |\int V \mathrm{d}y|$.
This potential can be used to reproduce low-energy properties 
of two impurities when higher order terms in Eq.~(\ref{eq:expansion}) 
are not important. The parameter $\kappa$ can be obtained, for example,
from a single-phonon exchange~\cite{schecter2014},
in which case $\kappa=2 c^2/\pi^2\simeq 0.202 c^2$ for $c\to 0$. 
If the potential 
\begin{equation}
W_{ZR}(y_1-y_2)=\left[2c-2 c^2/\pi^2\right]\delta(y_1-y_2)
\end{equation}
is used in Eq.~(\ref{eq:ham_eff}) 
then a single bound state with the energy 
\begin{equation}
E_{\mathrm{eff}}^{ZR}=-\frac{m_{\mathrm{eff}}}{2m}(c-c^2/\pi^2)^2
\label{eq:Eeff_ZR}
\end{equation} 
is produced.
This effective model captures qualitatively the exact results, see Fig.~\ref{fig:Finalresult}~(top).
We show the ground state energies of $h_{\mathrm{eff}}$ with $m_{\mathrm{eff}}$ from Eq.~(\ref{eq:mass_renorm}) 
and with $m_{\mathrm{eff}}=m$, to
illustrate that the mass renormalization leads only to a marginal correction for 
the considered values of $c$.

{\it Induced interaction from the Born-Openheimer approximation.} 
The potential $V$ in Eq.~(\ref{eq:effpot}) can also be derived in the 
Born-Oppenheimer
approximation, where it is assumed that the Fermi gas follows the impurity adiabatically.
 The potential in this case is simply the energy 
in Fig.~\ref{fig:potentialAndIntegralPlot}.
This approximation must be accurate if the impurities are 
very heavy. For mobile impurities this approximation is useful 
if the impurities move slowly in comparison to the Fermi velocity, which defines 
the dispersion of a sound mode in a Fermi gas.

The function $E_{BO}(y_1-y_2)$ decays as $1/|y_1-y_2|$, however, it leads to an effectively 
short-range potential due to the oscillatory tail. For example, $\int E_{BO}(y)\mathrm{d}y$ is well-defined.
We calculate that $|\int E_{BO}(y)\mathrm{d}y|\simeq 0.22 c^2$ for $c\to 0$.
This is slightly larger than $\kappa \simeq 0.202c^2$ for the zero-range
potential discussed above. Even though, the long-range tail is not expected for integrable systems~\cite{schecter2014}, the potential $E_{BO}$ performs as well as the zero-range potential, confirming that only integral properties of $V$ are important.
Figure~\ref{fig:Finalresult}~(center) gives the binding energy calculated 
using the potential
\begin{equation}
W_{BO}=2c\delta(y_1-y_2)+E_{BO}(y_1-y_2).
\end{equation}
Figure~\ref{fig:Finalresult}~(bottom) shows the quantity $E+c^2/2$ to single out the effect of the induced interaction. The center and bottom panels of Fig.~\ref{fig:Finalresult} demonstrate that the $E_{BO}$ can be used to qualitatively analyze in-medium bound states.  
We note that the difference between the filled and empty circles in Fig.~\ref{fig:Finalresult}~(bottom) for $|c|<0.5$ can be fit with $-0.036c^2-0.1873 c^3$, which demonstrates the importance of terms with $c^n$, $n>2$.
To reduce this disagreement between the exact results and the effective model one can include coulpings between eigenstates 
of $h_{BO}$ due to the motion of the impurities. We leave this discussion for future studies.

To calculate the data in Fig.~\ref{fig:Finalresult}~(center, bottom), 
we exactly diagonalize the Hamiltonian $h_{\mathrm{eff}}$ from Eq.~(\ref{eq:ham_eff}) with $W_{BO}$: We use the eigenstates of the effective Hamiltonian with $W_{BO}=0$
as a basis to write $h_{\mathrm{eff}}$ as a matrix and diagonalize this matrix after truncation. 
The energy is found by fitting to the form $E_{K}=E_{\mathrm{eff}}+D/K$,
where $K$ is the number of used basis states and $D, E_{\mathrm{eff}}$ 
are fitting parameters. This slow $1/K$-convergence  is expected 
for zero-range interactions~\cite{Volosniev_2017a}.

{\it Other potentials.} The effective potential can also be calculated using other approximation schemes.
For example, trial wave functions~\cite{adamovski1989, Vansant_1994, volosniev2018}
can be used. We do not discuss those approaches here. However, we note that 
different methods may lead to different shapes of the effective potential.
This is not surprising, since the effective potential is not an observable 
quantity for mobile impurities.

\subsection{Non-integrable cases}
\label{sec:non_int}

\begin{figure}[htb]
\centering
\includegraphics[width=0.9\linewidth]{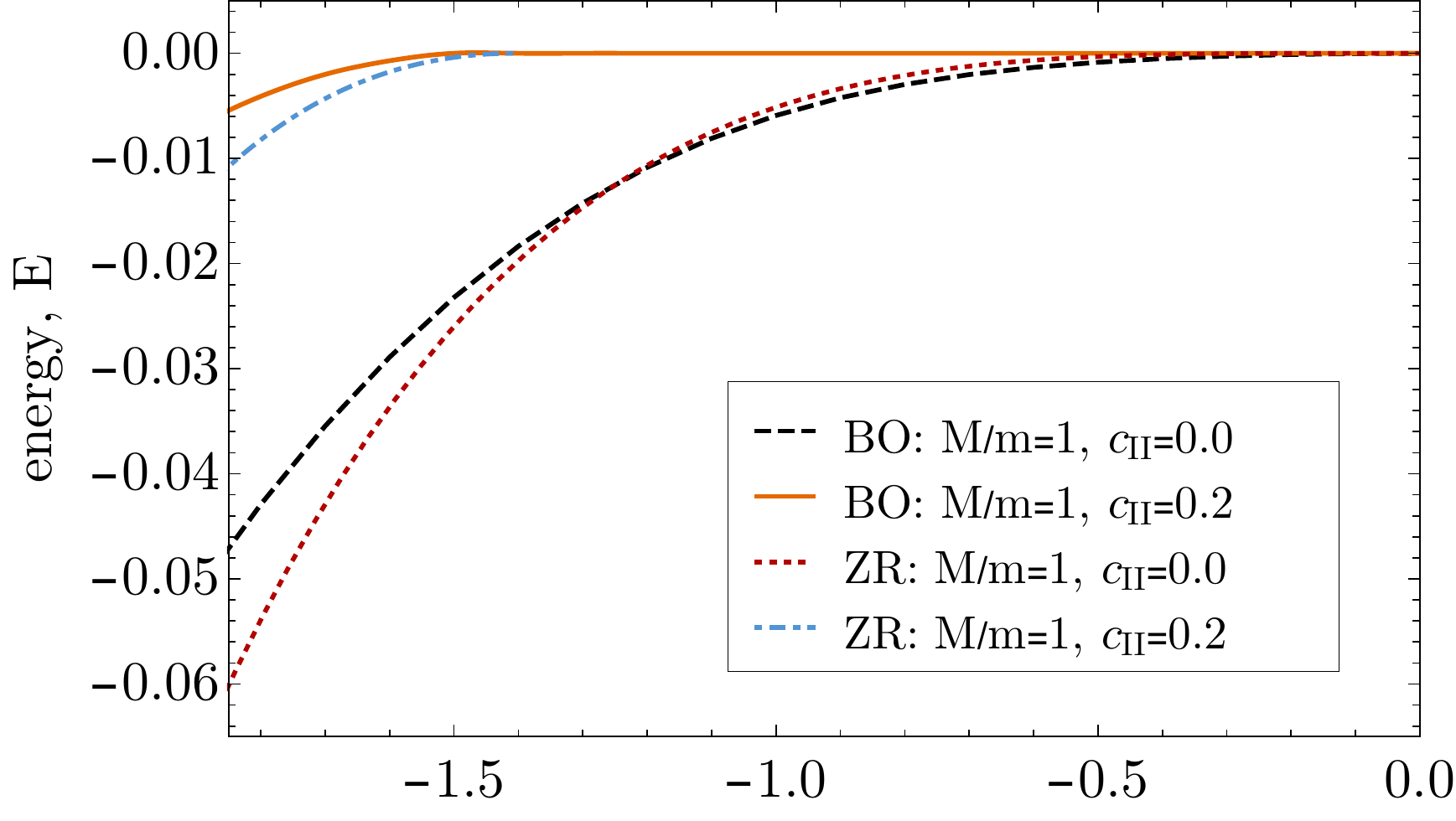} 
\includegraphics[width=0.9\linewidth]{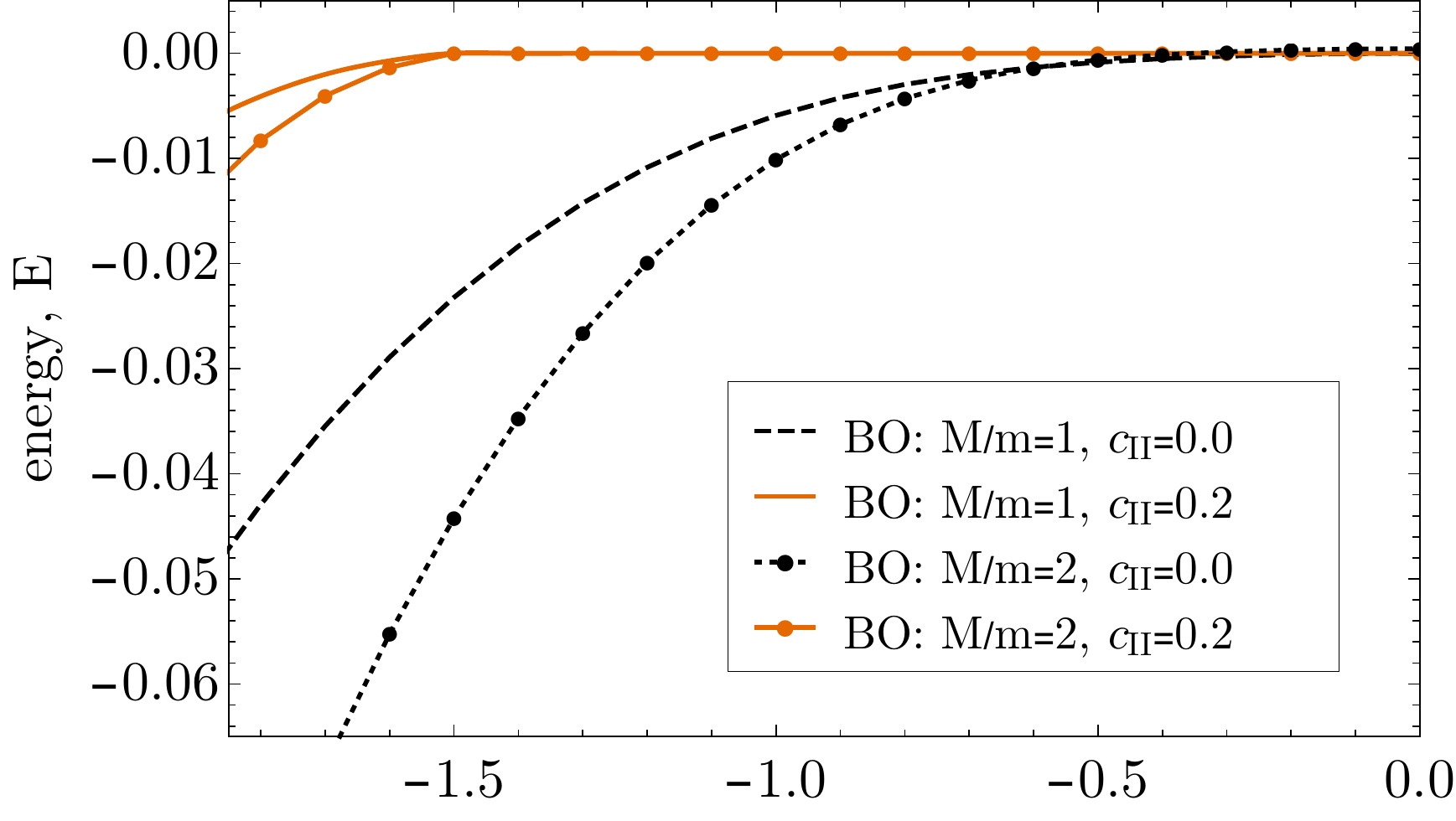}
\includegraphics[width=0.9\linewidth]{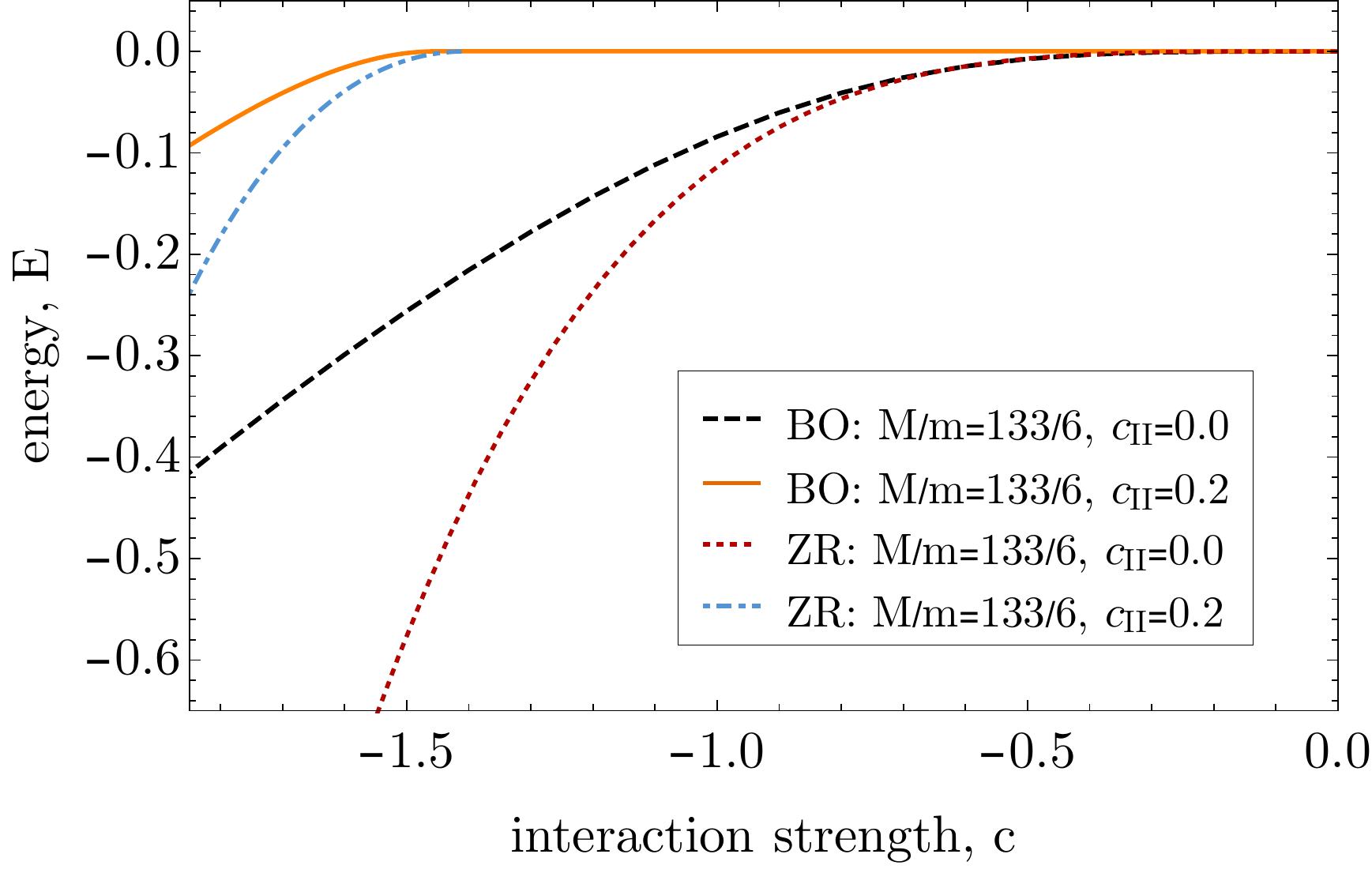} 
\caption{
The ground-state energy for two impurities in a Fermi gas in
non-integrable cases. We show results for the effective Hamiltonian
$h_{\mathrm{eff}}$ with $m_{\mathrm{eff}}=M$ for the BO potential from
Fig.~\ref{fig:potentialAndIntegralPlot} and the ZR
approximation in Eq.~(\ref{eq:Eeff_ZR}).
Top: comparison of ZR and BO results in the case $M=m$
for $c_{II}=0$ (two lower curves) and $c_{II}=0.2$ (two upper curves).
Middle: impact of the mass ratio. Comparison of the BO results for
$M=m$ (curves) and $M=2m$ (curves with dots) at $c_{II}=0$ and $c_{II}=0.2$.
Bottom: comparison of ZR and BO
potentials for the mass ratio of a $^{133}$Cs/$^6$Li mixture
at $c_{II}=0$ (two lower curves) and $c_{II}=0.2$ (two upper curves).
}
\label{fig:MassImBal}
\end{figure}

Motivated by the accuracy of the effective model~(\ref{eq:ham_eff})
for the most symmetric case, we extend this model to study appearance of in-medium bound states in non-integrable systems, i.e., $c_{II}\neq c$ and/or $m\neq M$. We write the corresponding effective Hamiltonian as
\begin{equation}
h_{\mathrm{eff}}=-\frac{m}{M}\frac{\partial^2}{\partial y_1^2}-\frac{m}{M}\frac{\partial^2}{\partial y_2^2}+2c_{II}\delta(y_1-y_2)+V(y_1-y_2),
\end{equation}
where we use $m_{\mathrm{eff}}=M$ for simplicity. This approximation relies on the observation that the mass renormalization is not important for the Bethe-ansatz-integrable case for weak interactions.
 $h_{\mathrm{eff}}$ supports a bound state for all $c_{II}<0$ because the induced interaction is attractive. For $c_{II}>0$ the bound states appear only if $c_{II}<c_{II}^{cr}$ for which
the repulsive impurity-impurity interaction in the absence of fermions is overtaken by the attractive interaction mediated by the Fermi gas. 

We first consider the case when $c_{II}\neq c$ and $m=M$, in which the particle exchange symmetry is broken 
by the interaction term~\footnote{For strong interactions, this case can still be studied with integrable Heisenberg Hamiltonians
using the mapping onto a spin chain~\cite{volosniev2014_nat,santos2014,volosniev2015,massignan2015}.
We do not discuss this limit here, instead we focus on the regimes with weak and moderate interactions}. 
The kinetic energy is still symmetric with respect to the exchange of two particles.
We calculate binding energies using the zero-range effective potential and the potential from the adiabatic approximation, see Fig.~\ref{fig:MassImBal}~(top). Both potentials lead to similar results for $c_{II}=0$ and small values of $c$. Note that in this regime the one-phonon exchange potential gives a leading contribution in $c$. For larger values of $c$ the ZR potential must be corrected. 

For $c_{II}=0.2$ the ZR potential predicts that the in-medium bound state is formed at smaller values of $c$  in comparison to the BO potential, see Fig.~\ref{fig:MassImBal}~(top). The difference between $c_{II}^{cr}$ in the two methods is, however, marginal and one can use
the ZR potential to derive the critical value for the appearance of the bound state:
\begin{equation}
c^{cr}_{II}\simeq \frac{c^2}{\pi^2}.
\label{eq:c_cr}
\end{equation}
The transition from an overall repulsive to an overall attractive induced interaction can 
be studied in Bose-Fermi mixtures by looking, for example,
at the collapse dynamics~\cite{chin2019}: a Bose gas can be stable only if its particles repel each other. One could also study spectroscopically the energy needed 
to break a bound state by transfering the system into a non-interacting state. 
Let us estimate the in-medium binding energy for $m=M$.  
For $c=-2$ and $c_{II}=0$, we have $E\simeq-0.06$ (see Fig.~\ref{fig:MassImBal}~(center)),
which for $^6$Li atoms with $\rho=3/(\mu m)$ translates into $\simeq 22 $ nK $\times k_{B}$, 
where $k_B$ is the Boltzmann constant. This means that in-medium bound states can be observed 
only at ultracold temperatures.

The in-medium binding energy can be increased if the impurities are heavy, cf.~Eq.~(\ref{eq:Eeff_weak}) with $m_{\mathrm{eff}}=M$.
To explicitly show this, we consider $c_{II}\neq c$ and $m\neq M$. In this case both the interaction 
and kinetic energies are not symmetric with respect to the exchange of two particles.
For the sake of discussion, we first use the BO potential to calculate the energies of the system with $M=2m$, see Fig.~\ref{fig:MassImBal}~(center).
The behavior of the energy resembles that for $m=M$, but, as expected, the overall energy scale 
is now larger. It is worthwhile noting that the critical value $c^{cr}_{II}$ obtained using 
the ZR approximation does not depend on the mass $M$. The dependence of $c^{cr}_{II}$ on $M$ is also negligible for the BO potential. Therefore, Eq.~(\ref{eq:c_cr}) can be used to predict the appearance of bound states for different masses of the impurity.

Finally, we consider two bosonic $^{133}$Cs atoms in a fermionic gas of $^6$Li as in the experiment of Ref.~\cite{chin2019}, see Fig.~\ref{fig:MassImBal}~(bottom). 
We use both the Born-Oppenheimer potential as well as the zero-range potential. Note however that for $M \gg m$  the former is expected to perform better than the latter. 
For a Li-Cs  mixture the energy scale is larger than that for lighter impurities and the bound states should be observable at much higher temperatures.

\section{Summary/Outlook}
\label{sec:summ}
We investigate the problem of two bosonic impurities in a spin-polarized
Fermi gas. First, we consider the ground state energy of the system in the
Bethe-ansatz-solvable case, i.e., equal masses of fermions and
impurities, $m=M$, as well as equal impurity-impurity and
impurity-fermion interactions, $c_{II}=c$. We
calculate the ground state energy and show that there are
attractive impurity-impurity interactions induced by the fermionic
medium. In the next step, we discuss an effective model for the
induced interactions and compare its predictions to the exact results.
We use two effective potentials to define the effective Hamiltonian
for the impurity system: a zero-range potential matched to single-phonon
exchange and an adiabatic potential for heavy impurities 
derived in the Born-Oppenheimer approximation. Both potentials
are able to approximate the exact results from the Bethe ansatz.
The difference between the two model potentials for $m=M$ allows us to estimate
the errors and the breakdown of our effective Hamiltonian.
For the Bethe-ansatz-integrable case the difference between the results derived using the two effective potentials is
marginal, which argues in favor of using them for qualitative analysis
of Fermi gases with impurities.

The success of the effective model in the integrable case motivates our use of the effective model to study non-integrable 
systems characterized by relaxing at least one of the two conditions
$m=M$ and  $c_{II}=c$. 
For repulsive impurity-impurity interactions without the Fermi gas, $c_{II}>0$,
we predict that the induced interaction overcomes the repulsion 
if the impurity-fermion interaction satisfies $c< -\pi\sqrt{c_{II}}$,
leading to an in-medium bound state.
The binding energies are larger for heavier impurities such that
the observation of in-medium bound states in heavy-light mixtures
appears more promising.

Our findings show that the Bethe-ansatz-solvable models provide a playground 
for investigating induced interactions. 
In the future it will be interesting to use the Bethe ansatz equations~(\ref{eq:BA}) 
to investigate spatial correlations of two impurities, which will 
allow us to test an effective model beyond the simple energy comparison
presented here. Further studies of non-integrable systems are also needed. The non-integrability 
due to $c_{II}\neq c$ and $m\neq M$ has been briefly discussed here.  
For cold atoms it is important also to consider trap effects,
which break the integrability and change the properties of the 
system~\cite{zinner2016, sowinski2019}. 

It will be interesting to extend the present study to fermionic impurities. 
It is known that two fermionic impurities in the SU(2) 
case do not have an in-medium bound state~\cite{lieb1967a}.
However, if the impurities are very heavy then a bound state must exist: 
The BO potential in Fig.~\ref{fig:potentialAndIntegralPlot}
has attractive regions and
unlike the zero-range potential 
of Eq.~(\ref{eq:zero_range_pot}) has a finite range.
Since it becomes exact as $M\to\infty$, the system must support a
bound state in this limit. This prediction can be explored using
numerical many-body methods
that are able to deal with mass-imbalanced systems such as 
the complex Langevin approach~\cite{Rammelmuller2017,Rammelm_ller_2018}.

Finally, we note that there is still a limited number of exact numerical results on two impurities   
in Bose gases. The present work gives 
insight into the properties of impurities in strongly-interacting Bose gases~\cite{Girardeau1960}. 
However, further work should be done to understand impurities in weakly-interacting Bose gases.
Since these systems cannot be studied using the Bethe ansatz, one has to employ other 
methods~\cite{Grusdt_2017, Pastukhov2017, volosniev2018, Volosniev_2017a, parisi2017, kain2018,mistakidis2019_quench,Paeckel:2019yjf,RevModPhys.77.259,RevModPhys.83.349, orso2019}.

\acknowledgements{We thank Lukas Rammelm{\"u}ller and Simos Mistakidis for useful discussions
and comments on the manuscript.
This  work
has been supported by the Deutsche Forschungsgemeinschaft
(DFG, German Research Foundation) under project numbers
413495248 -- VO 2437/1-1 and 279384907 -- SFB 1245, by the Bundesministerium 
f{\"u}r Bildung und Forschung (BMBF) through contract 05P18RDFN1,
 and by the European Union's Horizon 2020 research and innovation programme 
under the Marie Sk\l{}odowska-Curie Grant Agreement No. 754411.}


\begin{appendix}


\section{Weak coupling expansion}
\label{app:a}
In this appendix, we derive a weak coupling expansion of the BA equations,
\begin{align}
\label{eq:BAappendix1}
& e^{ik_jl}=\frac{k_j-\Lambda_1+\frac{ic}{2}}{k_j-\Lambda_1-\frac{ic}{2}} 
\frac{k_j-\Lambda_2+\frac{ic}{2}}{k_j-\Lambda_2-\frac{ic}{2}}\,,\quad 1\leq j\leq N;\\
\label{eq:BAappendix2}
& \prod_{j=1}^N\frac{k_j-\Lambda_1+\frac{ic}{2}}{k_j-\Lambda_1-\frac{ic}{2}}=1\,,\quad 
\prod_{j=1}^N\frac{k_j-\Lambda_2+\frac{ic}{2}}{k_j-\Lambda_2-\frac{ic}{2}}=1\,,
\end{align}
where $N$ is an odd number.
First, we consider the quasi-momenta $k_j$ ($j=3,...,N$) that satisfy $k_j(c=0)\neq 0$. For $c=0$, these quasi-momenta are multiples of $2\pi/l$. When $c\neq 0$, we write them as 
\begin{equation}
k_j=\frac{2\pi}{l}m_j+\delta_j\,, \quad m_j\in\mathbb{Z}\setminus\{0\}\,;
\label{eq:exp}
\end{equation}
where $\delta_j$ is small. Inserting Eq.~(\ref{eq:exp}) into the left-hand-side of Eq.~\eqref{eq:BAappendix1} leads to
\begin{equation}
\label{eq:LeftBA}
e^{ik_jl}= \underbrace{e^{im_j2\pi}}_{=1}e^{i\delta_jl}\approx 1+i\delta_jl\,.
\end{equation}
 We write the right-hand-side of Eq.~\eqref{eq:BAappendix1} as
\begin{align}
\label{eq:RightBa}
\begin{split}
\frac{k_j-\Lambda_1+\frac{ic}{2}}{k_j-\Lambda_1-\frac{ic}{2}}\frac{k_j-\Lambda_2+\frac{ic}{2}}{k_j-\Lambda_2-\frac{ic}{2}} \approx 1+\frac{2ic}{k_j^{(0)}}\,,
\end{split}
\end{align}
where terms proportional to $c^n$ with $n>1$ are neglected. Also it is assumed that $k_j^{(0)}\gg \delta_j-\Lambda_1$ and $k_j^{(0)}\gg \delta_j-\Lambda_2$.  This assumption is valid, since the $\Lambda$'s lie in between the first three quasi-momenta, which are all close to zero. 
To derive Eq.~(\ref{eq:RightBa}), we use that for $a\gg b$
\begin{align}
\label{eq:Approx}
\begin{split}
\frac{a+b}{a-b}&\approx 1+\frac{2b}{a}\,.
\end{split}
\end{align}
With Eqs.~\eqref{eq:LeftBA} and \eqref{eq:RightBa}, we obtain
\begin{equation}
\delta_jl=\frac{2c}{k_j^{(0)}}\quad\Rightarrow\quad \delta_j=\frac{c}{\pi m_j}\,.
\end{equation}

Now we investigate the quasi-momenta $k_j$ that vanish at $c=0$. As discussed in the main text, for the ground state, $k_1=-k_2$ and $k_3=0$.  To show that $\Lambda_1=-\Lambda_2$, we consider Eq.~\eqref{eq:BAappendix1} for $k_1$ and $k_2$. We use $k_1=-k_2$ in the equation for $k_2$:
\begin{align}
e^{-ik_1l}&=\frac{-k_1-\Lambda_1+\frac{ic}{2}}{-k_1-\Lambda_1-\frac{ic}{2}}\frac{-k_1-\Lambda_2+\frac{ic}{2}}{-k_1-\Lambda_2-\frac{ic}{2}}\,\\
\to \quad e^{ik_1l} &=\frac{k_1+\Lambda_1+\frac{ic}{2}}{k_1+\Lambda_1-\frac{ic}{2}}\frac{k_1+\Lambda_2+\frac{ic}{2}}{k_1+\Lambda_2-\frac{ic}{2}}\,.
\end{align}
From this equation and the equation for $k_1$:
\begin{equation}
e^{ik_1l}=\frac{k_1-\Lambda_1+\frac{ic}{2}}{k_1-\Lambda_1-\frac{ic}{2}}\frac{k_1-\Lambda_2+\frac{ic}{2}}{k_1-\Lambda_2-\frac{ic}{2}}\,,
\end{equation}
we obtain that $\Lambda_1=-\Lambda_2$. Using the equation for $k_1$, we derive
\begin{align}
\begin{split}
1+ik_1l \approx 1+\frac{2ick_1}{k_1^2-\Lambda_1^2}\\
\label{eq:ResOne}
\Leftrightarrow\quad(k_1^2-\Lambda_1^2)l & =2c\,.
\end{split}
\end{align}

Next, we consider Eq.~\eqref{eq:BAappendix2}. The sum of the quasi-momenta $\{k_3,...,k_N\}$ is zero, thus, Eq.~\eqref{eq:BAappendix2} can be rewritten as
\begin{equation}
\prod_{j=1}^{3}\frac{k_j-\Lambda_1+\frac{ic}{2}}{k_j-\Lambda_1-\frac{ic}{2}}=1\,.
\end{equation}
With $k_1=-k_2$, $k_3=0$, this equation reads
\begin{align}
\begin{split}
\frac{1}{k_1-\Lambda_1}-\frac{1}{k_1+\Lambda_1}-\frac{1}{\Lambda_1}=0\to \Lambda_1^2=\frac{k_1^2}{3}\,.
\end{split}
\end{align}
We rewrite Eq.~(\ref{eq:ResOne})
\begin{align}
(k_1^2-\Lambda_1^2)l=\frac{2}{3}k_1^2l=2c,
\end{align}
which leads to $k_{1}=\sqrt{3c/l}$ and $\Lambda_1=\sqrt{c/l}$.

\section{Numerical method to solve the BA equations}
\label{app:a0}

\begin{figure}
\centering
\includegraphics[width=\linewidth]{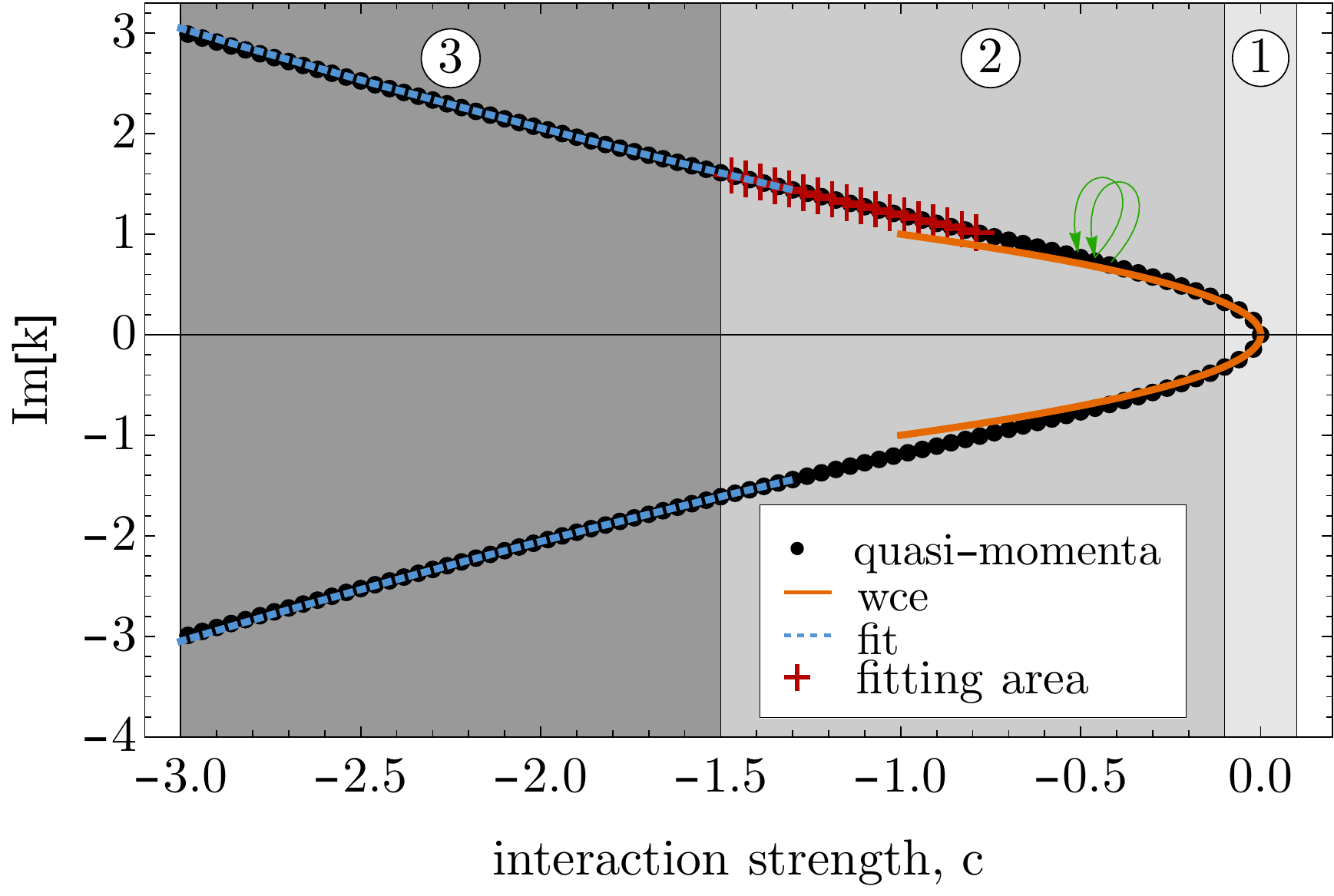} 
\caption{The black dots show the calculated quasi-momenta $k_1$ and $k_2$ [$k_1=-k_2$] as functions of the interaction strength, $c$. The area $1$ represents the region where the weak-coupling expansion (wce) is accurate. The (orange) solid curve is the weak-coupling expansion. The area $2$ is the region where we use the solution at the previous point (see the green arrows) as an initial guess for Newton's method. To establish an initial guess for the quasi-momenta in the area $3$, we construct a Taylor series using a number of already calculated points. The Taylor series constructed upon the points shown with crosses is presented as a (blue) dotted curve. }
\label{fig:app_B_2}
\end{figure}

\begin{figure}
\centering
\includegraphics[width=\linewidth]{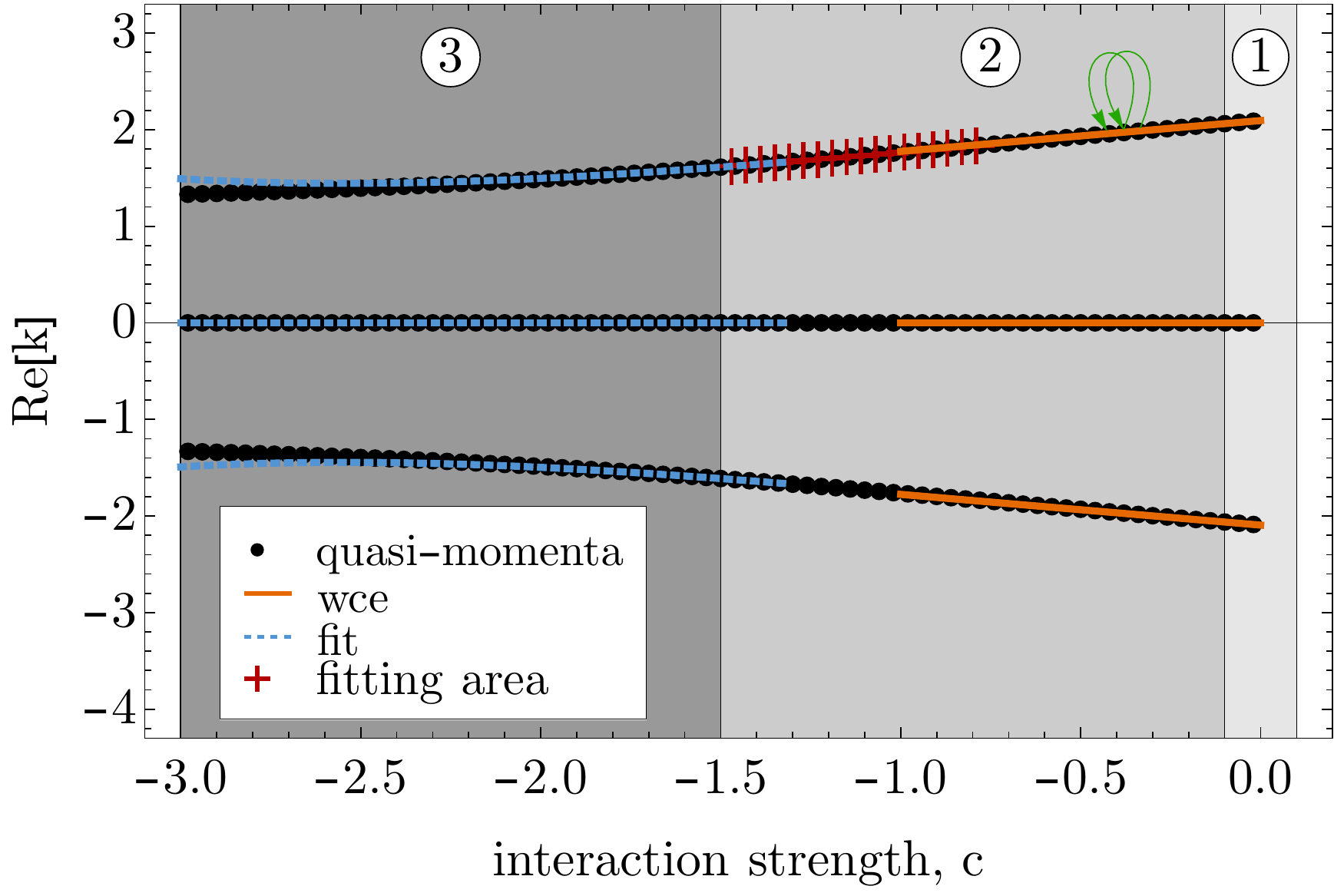} 
\caption{The black dots show the calculated quasi-momenta $k_3, k_4$ and $k_5$ [$k_3=0, k_5=-k_4$] as functions of the interaction strength, $c$. The area $1$ represents the region where the weak-coupling expansion (wce) is accurate. The (orange) solid curve is the weak-coupling expansion. The area $2$ is the region where we use a solution at the previous point (see the green arrows) as an initial guess for Newton's method. To establish an initial guess for the quasi-momenta in the area $3$, we construct a Taylor series using a number of already calculated points. The Taylor series constructed upon the points shown with crosses is presented as a (blue) dotted curve.     
  }
\label{fig:app_B_1}
\end{figure}

 In this appendix we give a detailed explanation of the numerical method that we employ to solve the BA equations. For simplicity, we consider here the case of 5 particles, three fermions and two bosons. To solve the BA equations, we employ Newton's method. Initial conditions for which are established differently for different regions of $c$. We exemplify different regions in Figs.~\ref{fig:app_B_2} and~\ref{fig:app_B_1}, which show the calculated values of $k_i$. The quasi-momenta $k_1$ and $k_2$ are purely imaginary, they are shown in  Fig.~\ref{fig:app_B_2}; the quasi-momenta $k_3,~k_4$ and $k_5$ are purely real, they are presented in Fig.~\ref{fig:app_B_1}. For better visibility, we do not show every value of $k_i$ that we calculate, only every fourth point. The three numbered areas with different types of shading refer to different methods we use to establish the initial conditions for Newton's method. 

For weak interactions (area 1) we use the weak coupling expansion from Eq.~(\ref{eq:initial_weak}), as an initial guess for Newton's method. As can be seen in Fig.~\ref{fig:app_B_2}, for $c\lesssim-0.5$ the deviation between Eq.~(\ref{eq:initial_weak}) and the exact solution becomes significant. We need another approach to calculate $k_i$ for stronger interactions.

In the second area, we use the solution at the previous step as an initial guess for Newton's method (see the green arrows in Fig.~\ref{fig:app_B_1}). At about $c\approx-1.5$, this method requires us to compute points lying very close to each other, which can be time consuming. So once more, we change our strategy.

To come up with an initial guess for Newton's method in the third region, we extrapolate the previously calculated solutions by using a polynomial fit with order 3. The red crosses in Fig.~\ref{fig:app_B_1}) exemplify a fitting range used to calculate the fitting function (dotted,blue curve). At $c\approx -2$ the fit function does not represent the exact solution well. Therefore, the fitting process must be repeated using the calculated solutions in the range $c\in (-2,-1.5)$ (not shown here for brevity).


\section{Thermodynamic extrapolation}
\label{app:b}

\begin{figure}
\centering
\includegraphics[width=\linewidth]{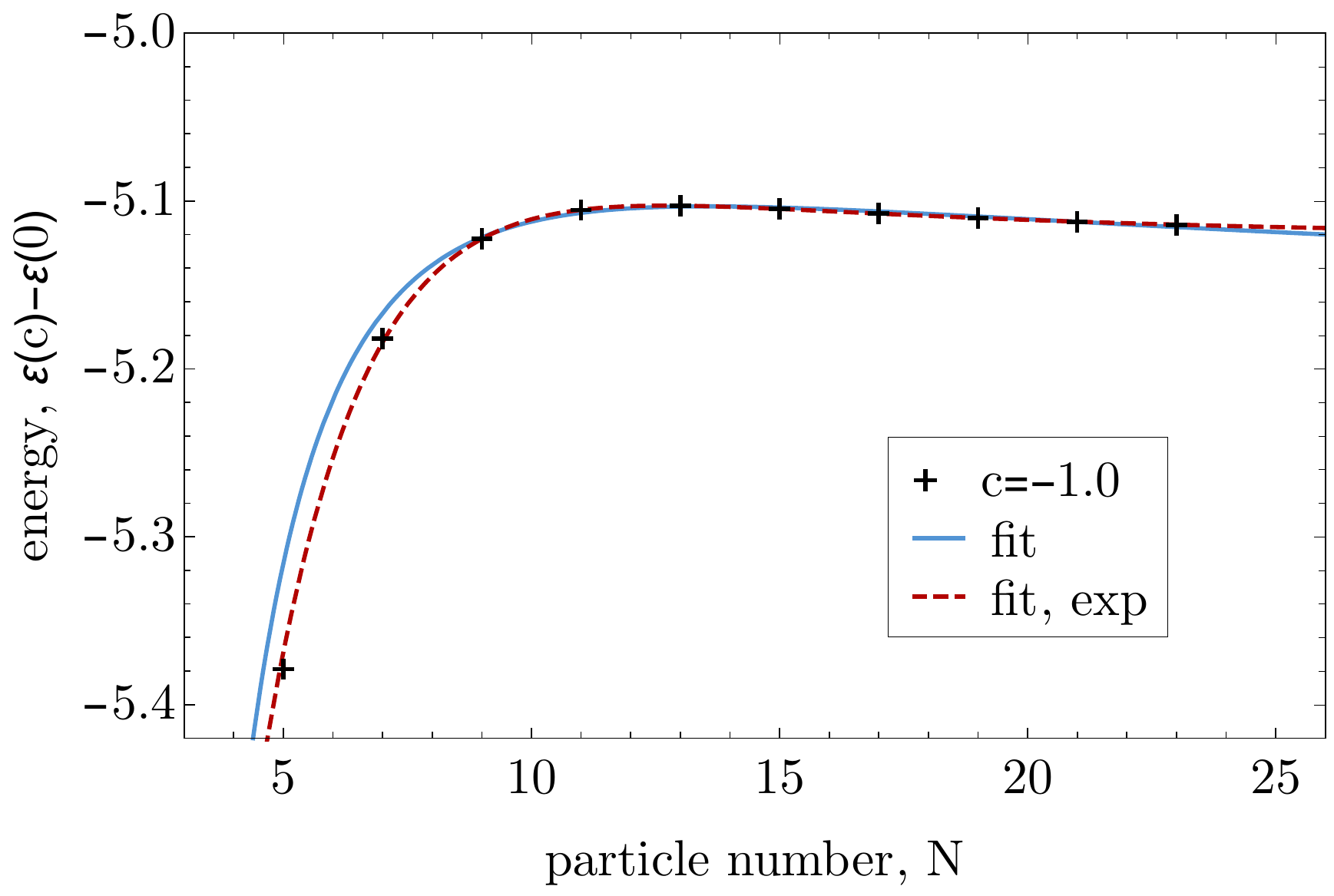} 
\caption{The energy $\varepsilon(c)-\varepsilon(0)$ as a function of the particle number $N$ for $c=-1.0$. The (blue) solid line corresponds to the fit with Eq.~\eqref{eq:fitfct1}, in which case $\varepsilon_\infty= -5.176$. The (red) dashed curve shows the fit with Eq.~\eqref{eq:fitfct2}, leading to $\varepsilon_\infty=-5.125$.}
\label{fig:TDextrapolationc1}
\end{figure}

\begin{figure}
\centering
\includegraphics[width=\linewidth]{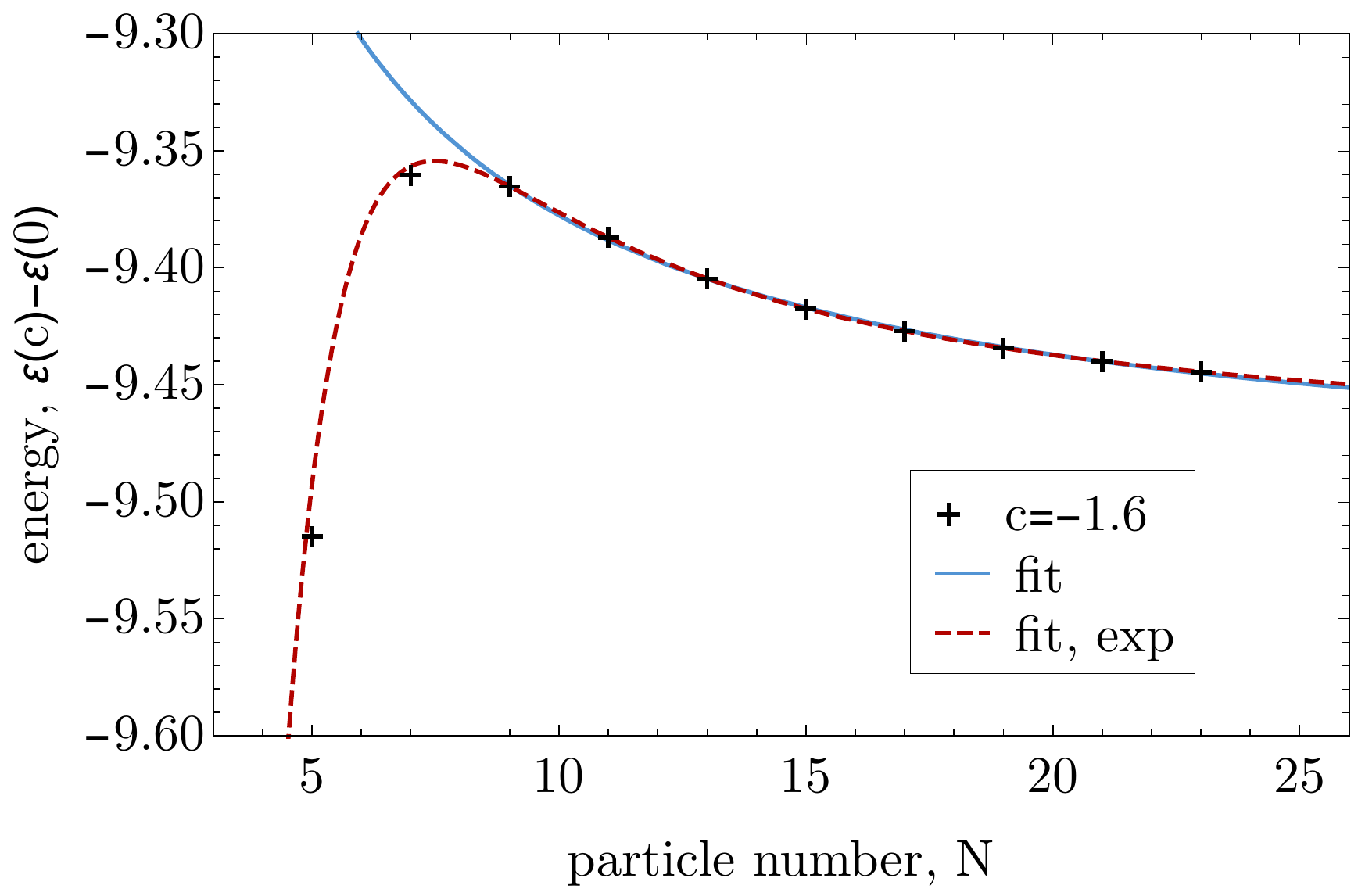} 
\caption{The energy $\varepsilon(c)-\varepsilon(0)$ as a function of the particle number $N$ for $c=-1.6$. The (blue) solid line corresponds to the fit with Eq.~\eqref{eq:fitfct1}, in which case $\varepsilon_\infty= -9.499$. The (red) dashed curve shows the fit with Eq.~\eqref{eq:fitfct2}, leading to $\varepsilon_\infty=-9.479$.}
\label{fig:TDextrapolationc2}
\end{figure}

To extrapolate the calculated energies $\varepsilon(c)-\varepsilon(0)$ to the thermodynamic limit, we shall employ the two fit functions:
\begin{align}
\label{eq:fitfct1}
& \text{1)}\quad \varepsilon_\infty+\frac{A_1}{N}+\frac{A_2}{N^2}\,,\\
\label{eq:fitfct2}
& \text{2)}\quad \varepsilon_\infty+\frac{A_1}{N^{\alpha}}+A_2\,e^{-\beta N}\,.
\end{align}
To illustrate the fits, we show in Figs.~\ref{fig:TDextrapolationc1} and \ref{fig:TDextrapolationc2} the exact energies as functions of $N$ for two different interaction strengths together with the corresponding fits. Both functions~(\ref{eq:fitfct1}) and~(\ref{eq:fitfct2}) appear to represent the data well. They also produce similar results for $N\to\infty$. The values of $\varepsilon_\infty$ from the two fits differ only in the third digit, implying that the precise knowledge of the convergence pattern as $N\to\infty$ is not needed for the considered parameters.

\section{One impurity}
\label{app:c}
Here we briefly review how to derive the ground state energy of a Fermi gas with a single impurity atom. This system has already been investigated~\cite{McGuire1965}. We use this well-known set-up to test our numerical approach. The system is desribed the Hamiltonian  
\begin{equation}
h_{BA}=-\sum_{j=1}^{N}\frac{\partial^2}{\partial x_j^2}+2c\sum_{j<l}^N\delta \left( x_j-x_l \right)\,,
\end{equation}
where the coordinates $x_1,...,x_{N-1}$ are the positions of the fermions, and $x_N$ is reserved for the impurity.
The corresponding Bethe ansatz equations are given by~\cite{yang1967}
\begin{align}
\begin{split}
e^{ik^{(1)}_jl}=&\frac{k^{(1)}_j-\Lambda+\frac{1}{2}ic}{k^{(1)}_j-\Lambda-\frac{1}{2}ic}\,,\quad 1\leq j\leq N; \\
\prod_{j=1}^N &\frac{k^{(1)}_j-\Lambda+\frac{1}{2}ic}{k^{(1)}_j-\Lambda-\frac{1}{2}ic}=1\,,
\end{split}
\end{align}
where $k^{(1)}_j$ is the $j$th quasi-momentum (we use the superscript to emphasize that we work with a single impurity here), and $\Lambda$ is one additional variable. We consider $N$ to be even. Once the BA equation are solved, the energy can be calculated as $\varepsilon^{(1)}=\sum_{j}\left(k^{(1)}_j\right)^2$.

We solve the BA equations with Newton's method as already explained in the main part. For small $c$ the weak coupling expansion of the BA equations~\cite{oelkers2006} is used as an initial guess
\begin{align}
\begin{split}
k^{(1)}_{1}&\approx-\frac{1}{4 \pi} \sum_{j=3}^N \frac{1}{m_j}c +  \sqrt{\frac{c}{l}}\,,\\
k^{(1)}_{2}&\approx-\frac{1}{4 \pi} \sum_{j=3}^N \frac{1}{m_j}c -  \sqrt{\frac{c}{l}}\,,\\
k^{(1)}_j&\approx\frac{2 \pi}{l}m_j + \frac{1}{2 \pi m_j} c\quad \text{for } 3\leq j\leq N\,;
\Lambda\approx \gamma c\,,
\end{split}
\end{align}
where $m_j \in \mathbb{Z}$ determine the quasi-momenta of the particles for zero interaction.  The shift due to the small interaction is given by the terms proportional to $c$ and $\sqrt{c}$.

\begin{figure}
\centering
\includegraphics[width=\linewidth]{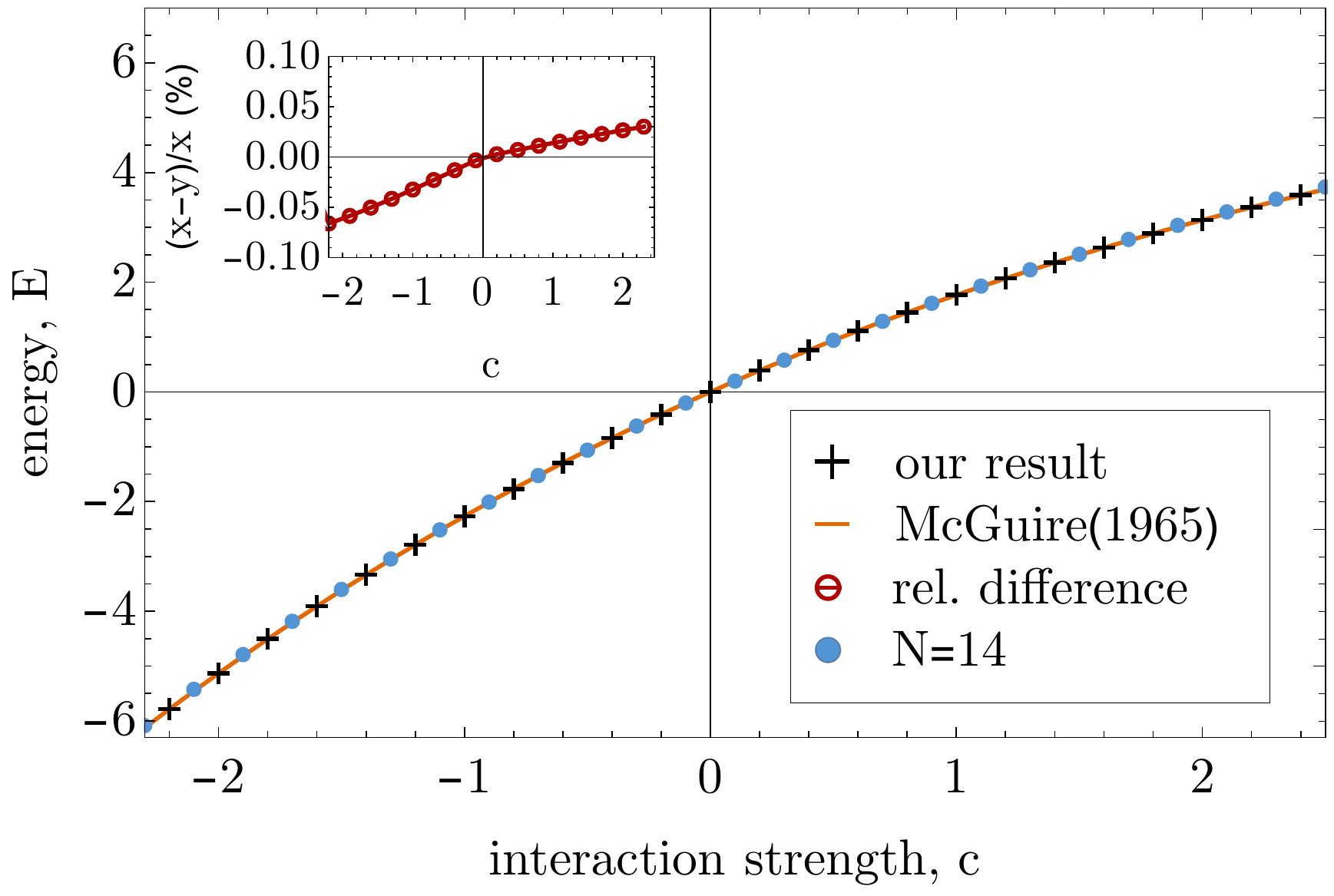} 
\caption{The energy of an impurity atom  in a Fermi gas, $\mathcal{E}$, as a function of the interaction strength, $c$. The crosses represent our numerical result for the thermodynamic limit. The analytic result for the thermodynamic limit~\cite{McGuire1965} is shown by the orange line.
In addition the total energy for a system consisting of $N=14$ particles is shown by the blue dots.
Inset: The red circles display the relative difference $(x-y)/x$, where $x$ is our numeric result and $y$ the analytic expression from~Ref.~\cite{McGuire1965}. The curve is added to guide the eye.}
\label{fig:TDlimitOneImpurity}
\end{figure}

To extrapolate the result to the thermodynamic limit we use $\varepsilon^{(1)}(c)-\varepsilon^{(1)}(0)=\mathcal{E}+\alpha/N^\beta$, where $\mathcal{E}, \alpha$ and $\beta$ are fit parameters. We show the result in Fig. \ref{fig:TDlimitOneImpurity}.
Our result fits the analytic expression quite well. The relative difference, shown in the inset, is negligible for our purposes. We present also the result for 14 particles to demonstrate that only a handful of particles are needed to simulate the ground state properties of an infinite Fermi gas with an impurity in a laboratory.

\section{Two static impurities}
\label{app:d}
Here we calculate the eigenenergies of the Hamiltonian
\begin{align}
h_{BO}=-\frac{\partial^2}{\partial x^2} + 2c\left[ \delta \left(x-\frac{r}{2} \right)+\delta \left(x+\frac{r}{2} \right)\right]\,.
\end{align}
To this end, we divide the space into three parts: $-l/2<x<-r/2$, $-r/2<x<r/2$, and $r/2<x<l/2$. For each part we write the wave function as
\begin{align}
\psi(x)=a_1 e^{i k x}+a_2 e^{-i k x}\,, 
\end{align}
where $k$ is the wave number. The wave function must obey the ``delta-potential boundary conditions'' at $x=\pm r/2$, and periodic boundary conditions at $x=\pm l$. 
We divide the solutions into parity-symmetric and parity-antisymmetric ones. The energy $k^2$ must be represented by a real number, which means that $k$ can be either real or imaginary. These two possibilities are referred to as the ``scattering states" and the ``bound states", respectively. For ``bound states", we write $k=i\kappa$, where $\kappa$ is real. The equations that determine $k$ for each class of solutions are written below. 

1) Symmetric ``bound states":
\begin{align}
\begin{split}
 -2 \kappa  e^{\frac{\kappa  l}{2}+\kappa  r}  & \left(  2c e^{-\kappa l  }+g e^{-\kappa l-\kappa  r}+2 c e^{-\kappa r}+2 c e^{-2 \kappa  r} \right. \\
& \left. -2 \kappa  e^{- \kappa l-\kappa  r}+2 \kappa  e^{-\kappa -r} \right) =0.
\end{split}
\end{align}
2) Antisymmetric ``bound states":
\begin{align}
\begin{split}
2 e^{\frac{\kappa  l}{2}+\kappa  r} & \left(2 c e^{-\kappa l}-2 c e^{-\kappa l-\kappa  r}-2 c e^{-\kappa r}+2 c e^{-2 \kappa  r} \right. \\
& \left. +2 \kappa  e^{-\kappa l-\kappa  r}-2 \kappa  e^{-\kappa r}\right)=0.
\end{split}
\end{align}
3) Symmetric ``scattering states":
\begin{align}
\begin{split}
k & \left[ 2 c  \left(\cos\left(\frac{kl}{2}\right)  + \cos\left(\frac{1}{2} k (l - 2 r)\right)\right) \right. \\
& \left. - 2 k \sin(\frac{kl}{2}) \right]=0.
\end{split}
\end{align}
4) Antisymmetric ``scattering states":
\begin{align}
8 c & \cos \left(\frac{1}{2} k (l-2 r)\right) -8 c \cos \left(\frac{k l}{2}\right) \\
& +8 k \sin \left(\frac{k l}{2}\right)=0 \,.
\end{align}
To solve the equations, a genetic algorithm first finds approximate solutions for $k, \kappa$. These are then used in the Newton's iteration method as an initial guess. We calculate as many energy levels as particles we consider.


\end{appendix}

\bibliographystyle{apsrev4-1}
\bibliography{bib}

\begin{thebibliography}{76}%
\makeatletter
\providecommand \@ifxundefined [1]{%
 \@ifx{#1\undefined}
}%
\providecommand \@ifnum [1]{%
 \ifnum #1\expandafter \@firstoftwo
 \else \expandafter \@secondoftwo
 \fi
}%
\providecommand \@ifx [1]{%
 \ifx #1\expandafter \@firstoftwo
 \else \expandafter \@secondoftwo
 \fi
}%
\providecommand \natexlab [1]{#1}%
\providecommand \enquote  [1]{``#1''}%
\providecommand \bibnamefont  [1]{#1}%
\providecommand \bibfnamefont [1]{#1}%
\providecommand \citenamefont [1]{#1}%
\providecommand \href@noop [0]{\@secondoftwo}%
\providecommand \href [0]{\begingroup \@sanitize@url \@href}%
\providecommand \@href[1]{\@@startlink{#1}\@@href}%
\providecommand \@@href[1]{\endgroup#1\@@endlink}%
\providecommand \@sanitize@url [0]{\catcode `\\12\catcode `\$12\catcode
  `\&12\catcode `\#12\catcode `\^12\catcode `\_12\catcode `\%12\relax}%
\providecommand \@@startlink[1]{}%
\providecommand \@@endlink[0]{}%
\providecommand \url  [0]{\begingroup\@sanitize@url \@url }%
\providecommand \@url [1]{\endgroup\@href {#1}{\urlprefix }}%
\providecommand \urlprefix  [0]{URL }%
\providecommand \Eprint [0]{\href }%
\providecommand \doibase [0]{http://dx.doi.org/}%
\providecommand \selectlanguage [0]{\@gobble}%
\providecommand \bibinfo  [0]{\@secondoftwo}%
\providecommand \bibfield  [0]{\@secondoftwo}%
\providecommand \translation [1]{[#1]}%
\providecommand \BibitemOpen [0]{}%
\providecommand \bibitemStop [0]{}%
\providecommand \bibitemNoStop [0]{.\EOS\space}%
\providecommand \EOS [0]{\spacefactor3000\relax}%
\providecommand \BibitemShut  [1]{\csname bibitem#1\endcsname}%
\let\auto@bib@innerbib\@empty
\bibitem [{\citenamefont {Landau}\ and\ \citenamefont
  {Pekar}(1948)}]{landau1948}%
  \BibitemOpen
  \bibfield  {author} {\bibinfo {author} {\bibfnamefont {L.~D.}\ \bibnamefont
  {Landau}}\ and\ \bibinfo {author} {\bibfnamefont {S.~I.}\ \bibnamefont
  {Pekar}},\ }\href@noop {} {\bibfield  {journal} {\bibinfo  {journal} {J. Exp.
  Theor. Phys}\ }\textbf {\bibinfo {volume} {18}},\ \bibinfo {pages} {419}
  (\bibinfo {year} {1948})}\BibitemShut {NoStop}%
\bibitem [{\citenamefont {Pekar.}(1963)}]{pekar1951}%
  \BibitemOpen
  \bibfield  {author} {\bibinfo {author} {\bibfnamefont {S.}~\bibnamefont
  {Pekar.}},\ }\href@noop {} {\emph {\bibinfo {title} {Research in Electron
  Theory of Crystals}}}\ (\bibinfo  {publisher} {AEC-tr-555, US Atomic Energy
  Commission},\ \bibinfo {year} {1963})\BibitemShut {NoStop}%
\bibitem [{\citenamefont {Alexandrov}\ and\ \citenamefont
  {Mott}(1995)}]{mott1995}%
  \BibitemOpen
  \bibfield  {author} {\bibinfo {author} {\bibfnamefont {A.~S.}\ \bibnamefont
  {Alexandrov}}\ and\ \bibinfo {author} {\bibfnamefont {N.~F.}\ \bibnamefont
  {Mott}},\ }\href@noop {} {\emph {\bibinfo {title} {Polarons and
  Bipolarons}}}\ (\bibinfo  {publisher} {World Scientific, Singapore},\
  \bibinfo {year} {1995})\BibitemShut {NoStop}%
\bibitem [{\citenamefont {Baym}\ and\ \citenamefont
  {Pethick}(2008)}]{baym2008}%
  \BibitemOpen
  \bibfield  {author} {\bibinfo {author} {\bibfnamefont {G.}~\bibnamefont
  {Baym}}\ and\ \bibinfo {author} {\bibfnamefont {C.}~\bibnamefont {Pethick}},\
  }\href@noop {} {\emph {\bibinfo {title} {Landau Fermi-Liquid Theory: Concepts
  and Applications}}}\ (\bibinfo {year} {2008})\BibitemShut {NoStop}%
\bibitem [{\citenamefont {Kutschera}\ and\ \citenamefont
  {W\'ojcik}(1993)}]{kutschera1993}%
  \BibitemOpen
  \bibfield  {author} {\bibinfo {author} {\bibfnamefont {M.}~\bibnamefont
  {Kutschera}}\ and\ \bibinfo {author} {\bibfnamefont {W.}~\bibnamefont
  {W\'ojcik}},\ }\href {\doibase 10.1103/PhysRevC.47.1077} {\bibfield
  {journal} {\bibinfo  {journal} {Phys. Rev. C}\ }\textbf {\bibinfo {volume}
  {47}},\ \bibinfo {pages} {1077} (\bibinfo {year} {1993})}\BibitemShut
  {NoStop}%
\bibitem [{\citenamefont {Schirotzek}\ \emph {et~al.}(2009)\citenamefont
  {Schirotzek}, \citenamefont {Wu}, \citenamefont {Sommer},\ and\ \citenamefont
  {Zwierlein}}]{zwierlein2009}%
  \BibitemOpen
  \bibfield  {author} {\bibinfo {author} {\bibfnamefont {A.}~\bibnamefont
  {Schirotzek}}, \bibinfo {author} {\bibfnamefont {C.-H.}\ \bibnamefont {Wu}},
  \bibinfo {author} {\bibfnamefont {A.}~\bibnamefont {Sommer}}, \ and\ \bibinfo
  {author} {\bibfnamefont {M.~W.}\ \bibnamefont {Zwierlein}},\ }\href {\doibase
  10.1103/PhysRevLett.102.230402} {\bibfield  {journal} {\bibinfo  {journal}
  {Phys. Rev. Lett.}\ }\textbf {\bibinfo {volume} {102}},\ \bibinfo {pages}
  {230402} (\bibinfo {year} {2009})}\BibitemShut {NoStop}%
\bibitem [{\citenamefont {Nascimb\`ene}\ \emph {et~al.}(2009)\citenamefont
  {Nascimb\`ene}, \citenamefont {Navon}, \citenamefont {Jiang}, \citenamefont
  {Tarruell}, \citenamefont {Teichmann}, \citenamefont {McKeever},
  \citenamefont {Chevy},\ and\ \citenamefont {Salomon}}]{salomon2009}%
  \BibitemOpen
  \bibfield  {author} {\bibinfo {author} {\bibfnamefont {S.}~\bibnamefont
  {Nascimb\`ene}}, \bibinfo {author} {\bibfnamefont {N.}~\bibnamefont {Navon}},
  \bibinfo {author} {\bibfnamefont {K.~J.}\ \bibnamefont {Jiang}}, \bibinfo
  {author} {\bibfnamefont {L.}~\bibnamefont {Tarruell}}, \bibinfo {author}
  {\bibfnamefont {M.}~\bibnamefont {Teichmann}}, \bibinfo {author}
  {\bibfnamefont {J.}~\bibnamefont {McKeever}}, \bibinfo {author}
  {\bibfnamefont {F.}~\bibnamefont {Chevy}}, \ and\ \bibinfo {author}
  {\bibfnamefont {C.}~\bibnamefont {Salomon}},\ }\href {\doibase
  10.1103/PhysRevLett.103.170402} {\bibfield  {journal} {\bibinfo  {journal}
  {Phys. Rev. Lett.}\ }\textbf {\bibinfo {volume} {103}},\ \bibinfo {pages}
  {170402} (\bibinfo {year} {2009})}\BibitemShut {NoStop}%
\bibitem [{\citenamefont {Massignan}\ \emph {et~al.}(2014)\citenamefont
  {Massignan}, \citenamefont {Zaccanti},\ and\ \citenamefont
  {Bruun}}]{massignan2014}%
  \BibitemOpen
  \bibfield  {author} {\bibinfo {author} {\bibfnamefont {P.}~\bibnamefont
  {Massignan}}, \bibinfo {author} {\bibfnamefont {M.}~\bibnamefont {Zaccanti}},
  \ and\ \bibinfo {author} {\bibfnamefont {G.~M.}\ \bibnamefont {Bruun}},\
  }\href@noop {} {\bibfield  {journal} {\bibinfo  {journal} {Rep. Prog. Phys.}\
  }\textbf {\bibinfo {volume} {77}},\ \bibinfo {pages} {034401} (\bibinfo
  {year} {2014})}\BibitemShut {NoStop}%
\bibitem [{\citenamefont {Hu}\ \emph {et~al.}(2016)\citenamefont {Hu},
  \citenamefont {Van~de Graaff}, \citenamefont {Kedar}, \citenamefont {Corson},
  \citenamefont {Cornell},\ and\ \citenamefont {Jin}}]{hu2016}%
  \BibitemOpen
  \bibfield  {author} {\bibinfo {author} {\bibfnamefont {M.-G.}\ \bibnamefont
  {Hu}}, \bibinfo {author} {\bibfnamefont {M.~J.}\ \bibnamefont {Van~de
  Graaff}}, \bibinfo {author} {\bibfnamefont {D.}~\bibnamefont {Kedar}},
  \bibinfo {author} {\bibfnamefont {J.~P.}\ \bibnamefont {Corson}}, \bibinfo
  {author} {\bibfnamefont {E.~A.}\ \bibnamefont {Cornell}}, \ and\ \bibinfo
  {author} {\bibfnamefont {D.~S.}\ \bibnamefont {Jin}},\ }\href {\doibase
  10.1103/PhysRevLett.117.055301} {\bibfield  {journal} {\bibinfo  {journal}
  {Phys. Rev. Lett.}\ }\textbf {\bibinfo {volume} {117}},\ \bibinfo {pages}
  {055301} (\bibinfo {year} {2016})}\BibitemShut {NoStop}%
\bibitem [{\citenamefont {J\o{}rgensen}\ \emph {et~al.}(2016)\citenamefont
  {J\o{}rgensen}, \citenamefont {Wacker}, \citenamefont {Skalmstang},
  \citenamefont {Parish}, \citenamefont {Levinsen}, \citenamefont
  {Christensen}, \citenamefont {Bruun},\ and\ \citenamefont {Arlt}}]{arlt2016}%
  \BibitemOpen
  \bibfield  {author} {\bibinfo {author} {\bibfnamefont {N.~B.}\ \bibnamefont
  {J\o{}rgensen}}, \bibinfo {author} {\bibfnamefont {L.}~\bibnamefont
  {Wacker}}, \bibinfo {author} {\bibfnamefont {K.~T.}\ \bibnamefont
  {Skalmstang}}, \bibinfo {author} {\bibfnamefont {M.~M.}\ \bibnamefont
  {Parish}}, \bibinfo {author} {\bibfnamefont {J.}~\bibnamefont {Levinsen}},
  \bibinfo {author} {\bibfnamefont {R.~S.}\ \bibnamefont {Christensen}},
  \bibinfo {author} {\bibfnamefont {G.~M.}\ \bibnamefont {Bruun}}, \ and\
  \bibinfo {author} {\bibfnamefont {J.~J.}\ \bibnamefont {Arlt}},\ }\href
  {\doibase 10.1103/PhysRevLett.117.055302} {\bibfield  {journal} {\bibinfo
  {journal} {Phys. Rev. Lett.}\ }\textbf {\bibinfo {volume} {117}},\ \bibinfo
  {pages} {055302} (\bibinfo {year} {2016})}\BibitemShut {NoStop}%
\bibitem [{\citenamefont {Schmidt}\ \emph {et~al.}(2018)\citenamefont
  {Schmidt}, \citenamefont {Knap}, \citenamefont {Ivanov}, \citenamefont {You},
  \citenamefont {Cetina},\ and\ \citenamefont {Demler}}]{schmidt2018}%
  \BibitemOpen
  \bibfield  {author} {\bibinfo {author} {\bibfnamefont {R.}~\bibnamefont
  {Schmidt}}, \bibinfo {author} {\bibfnamefont {M.}~\bibnamefont {Knap}},
  \bibinfo {author} {\bibfnamefont {D.~A.}\ \bibnamefont {Ivanov}}, \bibinfo
  {author} {\bibfnamefont {J.-S.}\ \bibnamefont {You}}, \bibinfo {author}
  {\bibfnamefont {M.}~\bibnamefont {Cetina}}, \ and\ \bibinfo {author}
  {\bibfnamefont {E.}~\bibnamefont {Demler}},\ }\href@noop {} {\bibfield
  {journal} {\bibinfo  {journal} {Rep. Prog. Phys.}\ }\textbf {\bibinfo
  {volume} {81}},\ \bibinfo {pages} {024401} (\bibinfo {year}
  {2018})}\BibitemShut {NoStop}%
\bibitem [{\citenamefont {Bruderer}\ \emph {et~al.}(2007)\citenamefont
  {Bruderer}, \citenamefont {Klein}, \citenamefont {Clark},\ and\ \citenamefont
  {Jaksch}}]{bruderer2007}%
  \BibitemOpen
  \bibfield  {author} {\bibinfo {author} {\bibfnamefont {M.}~\bibnamefont
  {Bruderer}}, \bibinfo {author} {\bibfnamefont {A.}~\bibnamefont {Klein}},
  \bibinfo {author} {\bibfnamefont {S.~R.}\ \bibnamefont {Clark}}, \ and\
  \bibinfo {author} {\bibfnamefont {D.}~\bibnamefont {Jaksch}},\ }\href
  {\doibase 10.1103/PhysRevA.76.011605} {\bibfield  {journal} {\bibinfo
  {journal} {Phys. Rev. A}\ }\textbf {\bibinfo {volume} {76}},\ \bibinfo
  {pages} {011605} (\bibinfo {year} {2007})}\BibitemShut {NoStop}%
\bibitem [{\citenamefont {Schecter}\ and\ \citenamefont
  {Kamenev}(2014)}]{schecter2014}%
  \BibitemOpen
  \bibfield  {author} {\bibinfo {author} {\bibfnamefont {M.}~\bibnamefont
  {Schecter}}\ and\ \bibinfo {author} {\bibfnamefont {A.}~\bibnamefont
  {Kamenev}},\ }\href {\doibase 10.1103/PhysRevLett.112.155301} {\bibfield
  {journal} {\bibinfo  {journal} {Phys. Rev. Lett.}\ }\textbf {\bibinfo
  {volume} {112}},\ \bibinfo {pages} {155301} (\bibinfo {year}
  {2014})}\BibitemShut {NoStop}%
\bibitem [{\citenamefont {Keiler}\ \emph {et~al.}(2018)\citenamefont {Keiler},
  \citenamefont {Krönke},\ and\ \citenamefont {Schmelcher}}]{Keiler_2018}%
  \BibitemOpen
  \bibfield  {author} {\bibinfo {author} {\bibfnamefont {K.}~\bibnamefont
  {Keiler}}, \bibinfo {author} {\bibfnamefont {S.}~\bibnamefont {Krönke}}, \
  and\ \bibinfo {author} {\bibfnamefont {P.}~\bibnamefont {Schmelcher}},\
  }\href {\doibase 10.1088/1367-2630/aab5e2} {\bibfield  {journal} {\bibinfo
  {journal} {New Journal of Physics}\ }\textbf {\bibinfo {volume} {20}},\
  \bibinfo {pages} {033030} (\bibinfo {year} {2018})}\BibitemShut {NoStop}%
\bibitem [{\citenamefont {Naidon}(2018)}]{naidon2018}%
  \BibitemOpen
  \bibfield  {author} {\bibinfo {author} {\bibfnamefont {P.}~\bibnamefont
  {Naidon}},\ }\href {\doibase 10.7566/JPSJ.87.043002} {\bibfield  {journal}
  {\bibinfo  {journal} {Journal of the Physical Society of Japan}\ }\textbf
  {\bibinfo {volume} {87}},\ \bibinfo {pages} {043002} (\bibinfo {year}
  {2018})}\BibitemShut {NoStop}%
\bibitem [{\citenamefont {Dehkharghani}\ \emph {et~al.}(2018)\citenamefont
  {Dehkharghani}, \citenamefont {Volosniev},\ and\ \citenamefont
  {Zinner}}]{volosniev2018}%
  \BibitemOpen
  \bibfield  {author} {\bibinfo {author} {\bibfnamefont {A.~S.}\ \bibnamefont
  {Dehkharghani}}, \bibinfo {author} {\bibfnamefont {A.~G.}\ \bibnamefont
  {Volosniev}}, \ and\ \bibinfo {author} {\bibfnamefont {N.~T.}\ \bibnamefont
  {Zinner}},\ }\href {\doibase 10.1103/PhysRevLett.121.080405} {\bibfield
  {journal} {\bibinfo  {journal} {Phys. Rev. Lett.}\ }\textbf {\bibinfo
  {volume} {121}},\ \bibinfo {pages} {080405} (\bibinfo {year}
  {2018})}\BibitemShut {NoStop}%
\bibitem [{\citenamefont {Camacho-Guardian}\ and\ \citenamefont
  {Bruun}(2018)}]{bruun2018a}%
  \BibitemOpen
  \bibfield  {author} {\bibinfo {author} {\bibfnamefont {A.}~\bibnamefont
  {Camacho-Guardian}}\ and\ \bibinfo {author} {\bibfnamefont {G.~M.}\
  \bibnamefont {Bruun}},\ }\href {\doibase 10.1103/PhysRevX.8.031042}
  {\bibfield  {journal} {\bibinfo  {journal} {Phys. Rev. X}\ }\textbf {\bibinfo
  {volume} {8}},\ \bibinfo {pages} {031042} (\bibinfo {year}
  {2018})}\BibitemShut {NoStop}%
\bibitem [{\citenamefont {Camacho-Guardian}\ \emph {et~al.}(2018)\citenamefont
  {Camacho-Guardian}, \citenamefont {Pe\~na Ardila}, \citenamefont {Pohl},\
  and\ \citenamefont {Bruun}}]{bruun2018}%
  \BibitemOpen
  \bibfield  {author} {\bibinfo {author} {\bibfnamefont {A.}~\bibnamefont
  {Camacho-Guardian}}, \bibinfo {author} {\bibfnamefont {L.~A.}\ \bibnamefont
  {Pe\~na Ardila}}, \bibinfo {author} {\bibfnamefont {T.}~\bibnamefont {Pohl}},
  \ and\ \bibinfo {author} {\bibfnamefont {G.~M.}\ \bibnamefont {Bruun}},\
  }\href {\doibase 10.1103/PhysRevLett.121.013401} {\bibfield  {journal}
  {\bibinfo  {journal} {Phys. Rev. Lett.}\ }\textbf {\bibinfo {volume} {121}},\
  \bibinfo {pages} {013401} (\bibinfo {year} {2018})}\BibitemShut {NoStop}%
\bibitem [{\citenamefont {Pavlov}\ \emph {et~al.}(2018)\citenamefont {Pavlov},
  \citenamefont {van~den Brink},\ and\ \citenamefont {Efremov}}]{pavlov2018}%
  \BibitemOpen
  \bibfield  {author} {\bibinfo {author} {\bibfnamefont {A.~I.}\ \bibnamefont
  {Pavlov}}, \bibinfo {author} {\bibfnamefont {J.}~\bibnamefont {van~den
  Brink}}, \ and\ \bibinfo {author} {\bibfnamefont {D.~V.}\ \bibnamefont
  {Efremov}},\ }\href {\doibase 10.1103/PhysRevB.98.161410} {\bibfield
  {journal} {\bibinfo  {journal} {Phys. Rev. B}\ }\textbf {\bibinfo {volume}
  {98}},\ \bibinfo {pages} {161410} (\bibinfo {year} {2018})}\BibitemShut
  {NoStop}%
\bibitem [{\citenamefont {Mistakidis}\ \emph
  {et~al.}(2019{\natexlab{a}})\citenamefont {Mistakidis}, \citenamefont
  {Katsimiga}, \citenamefont {Koutentakis},\ and\ \citenamefont
  {Schmelcher}}]{mistakidis2019}%
  \BibitemOpen
  \bibfield  {author} {\bibinfo {author} {\bibfnamefont {S.}~\bibnamefont
  {Mistakidis}}, \bibinfo {author} {\bibfnamefont {G.}~\bibnamefont
  {Katsimiga}}, \bibinfo {author} {\bibfnamefont {G.}~\bibnamefont
  {Koutentakis}}, \ and\ \bibinfo {author} {\bibfnamefont {P.}~\bibnamefont
  {Schmelcher}},\ }\href
  {https://iopscience.iop.org/article/10.1088/1367-2630/ab1045} {\bibfield
  {journal} {\bibinfo  {journal} {New Journal of Physics}\ }\textbf {\bibinfo
  {volume} {21}},\ \bibinfo {pages} {043032} (\bibinfo {year}
  {2019}{\natexlab{a}})}\BibitemShut {NoStop}%
\bibitem [{\citenamefont {Mistakidis}\ \emph
  {et~al.}(2019{\natexlab{b}})\citenamefont {Mistakidis}, \citenamefont
  {Hilbig},\ and\ \citenamefont {Schmelcher}}]{mistakidis2019_two}%
  \BibitemOpen
  \bibfield  {author} {\bibinfo {author} {\bibfnamefont {S.~I.}\ \bibnamefont
  {Mistakidis}}, \bibinfo {author} {\bibfnamefont {L.}~\bibnamefont {Hilbig}},
  \ and\ \bibinfo {author} {\bibfnamefont {P.}~\bibnamefont {Schmelcher}},\
  }\href {https://arxiv.org/abs/1905.02624} {\bibfield  {journal} {\bibinfo
  {journal} {{\bf arXiv}:1905.02624}\ } (\bibinfo {year}
  {2019}{\natexlab{b}})}\BibitemShut {NoStop}%
\bibitem [{\citenamefont {Reichert}\ \emph {et~al.}(2019)\citenamefont
  {Reichert}, \citenamefont {Ristivojevic},\ and\ \citenamefont
  {Petkovic}}]{petkovic2019}%
  \BibitemOpen
  \bibfield  {author} {\bibinfo {author} {\bibfnamefont {B.}~\bibnamefont
  {Reichert}}, \bibinfo {author} {\bibfnamefont {Z.}~\bibnamefont
  {Ristivojevic}}, \ and\ \bibinfo {author} {\bibfnamefont {A.}~\bibnamefont
  {Petkovic}},\ }\href
  {https://iopscience.iop.org/article/10.1088/1367-2630/ab1b8e} {\bibfield
  {journal} {\bibinfo  {journal} {New Journal of Physics}\ }\textbf {\bibinfo
  {volume} {21}},\ \bibinfo {pages} {053024} (\bibinfo {year}
  {2019})}\BibitemShut {NoStop}%
\bibitem [{\citenamefont {DeSalvo}\ \emph {et~al.}(2019)\citenamefont
  {DeSalvo}, \citenamefont {Patel}, \citenamefont {Cai},\ and\ \citenamefont
  {Chin}}]{chin2019}%
  \BibitemOpen
  \bibfield  {author} {\bibinfo {author} {\bibfnamefont {B.~J.}\ \bibnamefont
  {DeSalvo}}, \bibinfo {author} {\bibfnamefont {K.}~\bibnamefont {Patel}},
  \bibinfo {author} {\bibfnamefont {G.}~\bibnamefont {Cai}}, \ and\ \bibinfo
  {author} {\bibfnamefont {C.}~\bibnamefont {Chin}},\ }\href
  {https://doi.org/10.1038/s41586-019-1055-0} {\bibfield  {journal} {\bibinfo
  {journal} {Nature}\ }\textbf {\bibinfo {volume} {568}},\ \bibinfo {pages}
  {61} (\bibinfo {year} {2019})}\BibitemShut {NoStop}%
\bibitem [{\citenamefont {Alexandrov}\ and\ \citenamefont
  {Mott}(1994)}]{mott1994}%
  \BibitemOpen
  \bibfield  {author} {\bibinfo {author} {\bibfnamefont {A.}~\bibnamefont
  {Alexandrov}}\ and\ \bibinfo {author} {\bibfnamefont {N.~F.}\ \bibnamefont
  {Mott}},\ }\href@noop {} {\bibfield  {journal} {\bibinfo  {journal} {Rep.
  Prog. Phys.}\ }\textbf {\bibinfo {volume} {57}},\ \bibinfo {pages} {1197}
  (\bibinfo {year} {1994})}\BibitemShut {NoStop}%
\bibitem [{\citenamefont {Adamowski}(1989)}]{adamovski1989}%
  \BibitemOpen
  \bibfield  {author} {\bibinfo {author} {\bibfnamefont {J.}~\bibnamefont
  {Adamowski}},\ }\href {\doibase 10.1103/PhysRevB.39.3649} {\bibfield
  {journal} {\bibinfo  {journal} {Phys. Rev. B}\ }\textbf {\bibinfo {volume}
  {39}},\ \bibinfo {pages} {3649} (\bibinfo {year} {1989})}\BibitemShut
  {NoStop}%
\bibitem [{\citenamefont {Schmickler}\ \emph {et~al.}(2019)\citenamefont
  {Schmickler}, \citenamefont {Hammer},\ and\ \citenamefont
  {Volosniev}}]{schmickler2019}%
  \BibitemOpen
  \bibfield  {author} {\bibinfo {author} {\bibfnamefont {C.}~\bibnamefont
  {Schmickler}}, \bibinfo {author} {\bibfnamefont {H.-W.}\ \bibnamefont
  {Hammer}}, \ and\ \bibinfo {author} {\bibfnamefont {A.}~\bibnamefont
  {Volosniev}},\ }\href {\doibase
  https://doi.org/10.1016/j.physletb.2019.135016} {\bibfield  {journal}
  {\bibinfo  {journal} {Physics Letters B}\ }\textbf {\bibinfo {volume}
  {798}},\ \bibinfo {pages} {135016} (\bibinfo {year} {2019})}\BibitemShut
  {NoStop}%
\bibitem [{\citenamefont {Takada}(1982)}]{takada1982}%
  \BibitemOpen
  \bibfield  {author} {\bibinfo {author} {\bibfnamefont {Y.}~\bibnamefont
  {Takada}},\ }\href {\doibase 10.1103/PhysRevB.26.1223} {\bibfield  {journal}
  {\bibinfo  {journal} {Phys. Rev. B}\ }\textbf {\bibinfo {volume} {26}},\
  \bibinfo {pages} {1223} (\bibinfo {year} {1982})}\BibitemShut {NoStop}%
\bibitem [{\citenamefont {Landau}\ and\ \citenamefont
  {Lifschitz}(1977)}]{landaubook}%
  \BibitemOpen
  \bibfield  {author} {\bibinfo {author} {\bibfnamefont {L.~D.}\ \bibnamefont
  {Landau}}\ and\ \bibinfo {author} {\bibfnamefont {E.~M.}\ \bibnamefont
  {Lifschitz}},\ }\href@noop {} {\emph {\bibinfo {title} {Quantum Mechanics:
  Non-relativistic Theory (3rd edition)}}}\ (\bibinfo  {publisher} {Elsevier
  Butterworth-Heinemann},\ \bibinfo {year} {1977})\BibitemShut {NoStop}%
\bibitem [{\citenamefont {Friedel}(1958)}]{Friedel1958}%
  \BibitemOpen
  \bibfield  {author} {\bibinfo {author} {\bibfnamefont {J.}~\bibnamefont
  {Friedel}},\ }\href {\doibase 10.1007/BF02751483} {\bibfield  {journal}
  {\bibinfo  {journal} {Il Nuovo Cimento (1955-1965)}\ }\textbf {\bibinfo
  {volume} {7}},\ \bibinfo {pages} {287} (\bibinfo {year} {1958})}\BibitemShut
  {NoStop}%
\bibitem [{\citenamefont {Cazalilla}\ \emph {et~al.}(2011)\citenamefont
  {Cazalilla}, \citenamefont {Citro}, \citenamefont {Giamarchi}, \citenamefont
  {Orignac},\ and\ \citenamefont {Rigol}}]{rigol2011}%
  \BibitemOpen
  \bibfield  {author} {\bibinfo {author} {\bibfnamefont {M.~A.}\ \bibnamefont
  {Cazalilla}}, \bibinfo {author} {\bibfnamefont {R.}~\bibnamefont {Citro}},
  \bibinfo {author} {\bibfnamefont {T.}~\bibnamefont {Giamarchi}}, \bibinfo
  {author} {\bibfnamefont {E.}~\bibnamefont {Orignac}}, \ and\ \bibinfo
  {author} {\bibfnamefont {M.}~\bibnamefont {Rigol}},\ }\href {\doibase
  10.1103/RevModPhys.83.1405} {\bibfield  {journal} {\bibinfo  {journal} {Rev.
  Mod. Phys.}\ }\textbf {\bibinfo {volume} {83}},\ \bibinfo {pages} {1405}
  (\bibinfo {year} {2011})}\BibitemShut {NoStop}%
\bibitem [{\citenamefont {Guan}\ \emph {et~al.}(2013)\citenamefont {Guan},
  \citenamefont {Batchelor},\ and\ \citenamefont {Lee}}]{guan2013}%
  \BibitemOpen
  \bibfield  {author} {\bibinfo {author} {\bibfnamefont {X.-W.}\ \bibnamefont
  {Guan}}, \bibinfo {author} {\bibfnamefont {M.~T.}\ \bibnamefont {Batchelor}},
  \ and\ \bibinfo {author} {\bibfnamefont {C.}~\bibnamefont {Lee}},\ }\href
  {\doibase 10.1103/RevModPhys.85.1633} {\bibfield  {journal} {\bibinfo
  {journal} {Rev. Mod. Phys.}\ }\textbf {\bibinfo {volume} {85}},\ \bibinfo
  {pages} {1633} (\bibinfo {year} {2013})}\BibitemShut {NoStop}%
\bibitem [{\citenamefont {Batchelor}\ and\ \citenamefont
  {Foerster}(2016)}]{Batchelor_2016}%
  \BibitemOpen
  \bibfield  {author} {\bibinfo {author} {\bibfnamefont {M.~T.}\ \bibnamefont
  {Batchelor}}\ and\ \bibinfo {author} {\bibfnamefont {A.}~\bibnamefont
  {Foerster}},\ }\href {\doibase 10.1088/1751-8113/49/17/173001} {\bibfield
  {journal} {\bibinfo  {journal} {Journal of Physics A: Mathematical and
  Theoretical}\ }\textbf {\bibinfo {volume} {49}},\ \bibinfo {pages} {173001}
  (\bibinfo {year} {2016})}\BibitemShut {NoStop}%
\bibitem [{\citenamefont {Wenz}\ \emph {et~al.}(2013)\citenamefont {Wenz},
  \citenamefont {Z{\"u}rn}, \citenamefont {Murmann}, \citenamefont {Brouzos},
  \citenamefont {Lompe},\ and\ \citenamefont {Jochim}}]{wenz2013}%
  \BibitemOpen
  \bibfield  {author} {\bibinfo {author} {\bibfnamefont {A.~N.}\ \bibnamefont
  {Wenz}}, \bibinfo {author} {\bibfnamefont {G.}~\bibnamefont {Z{\"u}rn}},
  \bibinfo {author} {\bibfnamefont {S.}~\bibnamefont {Murmann}}, \bibinfo
  {author} {\bibfnamefont {I.}~\bibnamefont {Brouzos}}, \bibinfo {author}
  {\bibfnamefont {T.}~\bibnamefont {Lompe}}, \ and\ \bibinfo {author}
  {\bibfnamefont {S.}~\bibnamefont {Jochim}},\ }\href {\doibase
  10.1126/science.1240516} {\bibfield  {journal} {\bibinfo  {journal}
  {Science}\ }\textbf {\bibinfo {volume} {342}},\ \bibinfo {pages} {457}
  (\bibinfo {year} {2013})}\BibitemShut {NoStop}%
\bibitem [{\citenamefont {Zinner}(2016)}]{zinner2016}%
  \BibitemOpen
  \bibfield  {author} {\bibinfo {author} {\bibfnamefont {N.~T.}\ \bibnamefont
  {Zinner}},\ }\href {\doibase 10.1051/epjconf/201611301002} {\bibfield
  {journal} {\bibinfo  {journal} {EPJ Web of Conferences}\ }\textbf {\bibinfo
  {volume} {113}},\ \bibinfo {pages} {01002} (\bibinfo {year}
  {2016})}\BibitemShut {NoStop}%
\bibitem [{\citenamefont {Rammelm\"uller}\ \emph
  {et~al.}(2017{\natexlab{a}})\citenamefont {Rammelm\"uller}, \citenamefont
  {Porter}, \citenamefont {Braun},\ and\ \citenamefont
  {Drut}}]{Rammelmuller2017a}%
  \BibitemOpen
  \bibfield  {author} {\bibinfo {author} {\bibfnamefont {L.}~\bibnamefont
  {Rammelm\"uller}}, \bibinfo {author} {\bibfnamefont {W.~J.}\ \bibnamefont
  {Porter}}, \bibinfo {author} {\bibfnamefont {J.}~\bibnamefont {Braun}}, \
  and\ \bibinfo {author} {\bibfnamefont {J.~E.}\ \bibnamefont {Drut}},\ }\href
  {\doibase 10.1103/PhysRevA.96.033635} {\bibfield  {journal} {\bibinfo
  {journal} {Phys. Rev. A}\ }\textbf {\bibinfo {volume} {96}},\ \bibinfo
  {pages} {033635} (\bibinfo {year} {2017}{\natexlab{a}})}\BibitemShut
  {NoStop}%
\bibitem [{\citenamefont {Pagano}\ \emph {et~al.}(2014)\citenamefont {Pagano},
  \citenamefont {Mancini}, \citenamefont {Cappellini}, \citenamefont
  {Lombardi}, \citenamefont {Sch{\"a}fer}, \citenamefont {Hu}, \citenamefont
  {Liu}, \citenamefont {Catani}, \citenamefont {Sias}, \citenamefont
  {Inguscio},\ and\ \citenamefont {Fallani}}]{fallani2014}%
  \BibitemOpen
  \bibfield  {author} {\bibinfo {author} {\bibfnamefont {G.}~\bibnamefont
  {Pagano}}, \bibinfo {author} {\bibfnamefont {M.}~\bibnamefont {Mancini}},
  \bibinfo {author} {\bibfnamefont {G.}~\bibnamefont {Cappellini}}, \bibinfo
  {author} {\bibfnamefont {P.}~\bibnamefont {Lombardi}}, \bibinfo {author}
  {\bibfnamefont {F.}~\bibnamefont {Sch{\"a}fer}}, \bibinfo {author}
  {\bibfnamefont {H.}~\bibnamefont {Hu}}, \bibinfo {author} {\bibfnamefont
  {X.-J.}\ \bibnamefont {Liu}}, \bibinfo {author} {\bibfnamefont
  {J.}~\bibnamefont {Catani}}, \bibinfo {author} {\bibfnamefont
  {C.}~\bibnamefont {Sias}}, \bibinfo {author} {\bibfnamefont {M.}~\bibnamefont
  {Inguscio}}, \ and\ \bibinfo {author} {\bibfnamefont {L.}~\bibnamefont
  {Fallani}},\ }\href {\doibase 10.1038/nphys2878} {\bibfield  {journal}
  {\bibinfo  {journal} {Nature Physics}\ }\textbf {\bibinfo {volume} {10}},\
  \bibinfo {pages} {198} (\bibinfo {year} {2014})}\BibitemShut {NoStop}%
\bibitem [{\citenamefont {Ferrier-Barbut}\ \emph {et~al.}(2014)\citenamefont
  {Ferrier-Barbut}, \citenamefont {Delehaye}, \citenamefont {Laurent},
  \citenamefont {Grier}, \citenamefont {Pierce}, \citenamefont {Rem},
  \citenamefont {Chevy},\ and\ \citenamefont {Salomon}}]{Ferrier-Barbut1035}%
  \BibitemOpen
  \bibfield  {author} {\bibinfo {author} {\bibfnamefont {I.}~\bibnamefont
  {Ferrier-Barbut}}, \bibinfo {author} {\bibfnamefont {M.}~\bibnamefont
  {Delehaye}}, \bibinfo {author} {\bibfnamefont {S.}~\bibnamefont {Laurent}},
  \bibinfo {author} {\bibfnamefont {A.~T.}\ \bibnamefont {Grier}}, \bibinfo
  {author} {\bibfnamefont {M.}~\bibnamefont {Pierce}}, \bibinfo {author}
  {\bibfnamefont {B.~S.}\ \bibnamefont {Rem}}, \bibinfo {author} {\bibfnamefont
  {F.}~\bibnamefont {Chevy}}, \ and\ \bibinfo {author} {\bibfnamefont
  {C.}~\bibnamefont {Salomon}},\ }\href {\doibase 10.1126/science.1255380}
  {\bibfield  {journal} {\bibinfo  {journal} {Science}\ }\textbf {\bibinfo
  {volume} {345}},\ \bibinfo {pages} {1035} (\bibinfo {year} {2014})},\ \Eprint
  {http://arxiv.org/abs/https://science.sciencemag.org/content/345/6200/1035.full.pdf}
  {https://science.sciencemag.org/content/345/6200/1035.full.pdf} \BibitemShut
  {NoStop}%
\bibitem [{\citenamefont {Wu}\ \emph {et~al.}(2011)\citenamefont {Wu},
  \citenamefont {Santiago}, \citenamefont {Park}, \citenamefont {Ahmadi},\ and\
  \citenamefont {Zwierlein}}]{wu2011}%
  \BibitemOpen
  \bibfield  {author} {\bibinfo {author} {\bibfnamefont {C.-H.}\ \bibnamefont
  {Wu}}, \bibinfo {author} {\bibfnamefont {I.}~\bibnamefont {Santiago}},
  \bibinfo {author} {\bibfnamefont {J.~W.}\ \bibnamefont {Park}}, \bibinfo
  {author} {\bibfnamefont {P.}~\bibnamefont {Ahmadi}}, \ and\ \bibinfo {author}
  {\bibfnamefont {M.~W.}\ \bibnamefont {Zwierlein}},\ }\href {\doibase
  10.1103/PhysRevA.84.011601} {\bibfield  {journal} {\bibinfo  {journal} {Phys.
  Rev. A}\ }\textbf {\bibinfo {volume} {84}},\ \bibinfo {pages} {011601}
  (\bibinfo {year} {2011})}\BibitemShut {NoStop}%
\bibitem [{\citenamefont {Yang}(1967)}]{yang1967}%
  \BibitemOpen
  \bibfield  {author} {\bibinfo {author} {\bibfnamefont {C.~N.}\ \bibnamefont
  {Yang}},\ }\href {\doibase 10.1103/PhysRevLett.19.1312} {\bibfield  {journal}
  {\bibinfo  {journal} {Phys. Rev. Lett.}\ }\textbf {\bibinfo {volume} {19}},\
  \bibinfo {pages} {1312} (\bibinfo {year} {1967})}\BibitemShut {NoStop}%
\bibitem [{\citenamefont {Lai}\ and\ \citenamefont {Yang}(1971)}]{yang1971}%
  \BibitemOpen
  \bibfield  {author} {\bibinfo {author} {\bibfnamefont {C.~K.}\ \bibnamefont
  {Lai}}\ and\ \bibinfo {author} {\bibfnamefont {C.~N.}\ \bibnamefont {Yang}},\
  }\href {\doibase 10.1103/PhysRevA.3.393} {\bibfield  {journal} {\bibinfo
  {journal} {Phys. Rev. A}\ }\textbf {\bibinfo {volume} {3}},\ \bibinfo {pages}
  {393} (\bibinfo {year} {1971})}\BibitemShut {NoStop}%
\bibitem [{\citenamefont {Batchelor}\ \emph {et~al.}(2005)\citenamefont
  {Batchelor}, \citenamefont {Bortz}, \citenamefont {Guan},\ and\ \citenamefont
  {Oelkers}}]{guan2005}%
  \BibitemOpen
  \bibfield  {author} {\bibinfo {author} {\bibfnamefont {M.~T.}\ \bibnamefont
  {Batchelor}}, \bibinfo {author} {\bibfnamefont {M.}~\bibnamefont {Bortz}},
  \bibinfo {author} {\bibfnamefont {X.~W.}\ \bibnamefont {Guan}}, \ and\
  \bibinfo {author} {\bibfnamefont {N.}~\bibnamefont {Oelkers}},\ }\href
  {\doibase 10.1103/PhysRevA.72.061603} {\bibfield  {journal} {\bibinfo
  {journal} {Phys. Rev. A}\ }\textbf {\bibinfo {volume} {72}},\ \bibinfo
  {pages} {061603} (\bibinfo {year} {2005})}\BibitemShut {NoStop}%
\bibitem [{\citenamefont {Imambekov}\ and\ \citenamefont
  {Demler}(2006)}]{demler2006}%
  \BibitemOpen
  \bibfield  {author} {\bibinfo {author} {\bibfnamefont {A.}~\bibnamefont
  {Imambekov}}\ and\ \bibinfo {author} {\bibfnamefont {E.}~\bibnamefont
  {Demler}},\ }\href {\doibase 10.1103/PhysRevA.73.021602} {\bibfield
  {journal} {\bibinfo  {journal} {Phys. Rev. A}\ }\textbf {\bibinfo {volume}
  {73}},\ \bibinfo {pages} {021602} (\bibinfo {year} {2006})}\BibitemShut
  {NoStop}%
\bibitem [{\citenamefont {McGuire}(1965)}]{McGuire1965}%
  \BibitemOpen
  \bibfield  {author} {\bibinfo {author} {\bibfnamefont {J.~B.}\ \bibnamefont
  {McGuire}},\ }\href {\doibase 10.1063/1.1704291} {\bibfield  {journal}
  {\bibinfo  {journal} {Journal of Mathematical Physics}\ }\textbf {\bibinfo
  {volume} {6}},\ \bibinfo {pages} {432} (\bibinfo {year} {1965})}\BibitemShut
  {NoStop}%
\bibitem [{\citenamefont {McGuire}(1966)}]{McGuire1966}%
  \BibitemOpen
  \bibfield  {author} {\bibinfo {author} {\bibfnamefont {J.~B.}\ \bibnamefont
  {McGuire}},\ }\href {\doibase 10.1063/1.1704798} {\bibfield  {journal}
  {\bibinfo  {journal} {Journal of Mathematical Physics}\ }\textbf {\bibinfo
  {volume} {7}},\ \bibinfo {pages} {123} (\bibinfo {year} {1966})}\BibitemShut
  {NoStop}%
\bibitem [{\citenamefont {McGuire}(1964)}]{mcguire1964}%
  \BibitemOpen
  \bibfield  {author} {\bibinfo {author} {\bibfnamefont {J.~B.}\ \bibnamefont
  {McGuire}},\ }\href {\doibase 10.1063/1.1704156} {\bibfield  {journal}
  {\bibinfo  {journal} {Journal of Mathematical Physics}\ }\textbf {\bibinfo
  {volume} {5}},\ \bibinfo {pages} {622} (\bibinfo {year} {1964})},\ \Eprint
  {http://arxiv.org/abs/https://doi.org/10.1063/1.1704156}
  {https://doi.org/10.1063/1.1704156} \BibitemShut {NoStop}%
\bibitem [{\citenamefont {Simon}(1976)}]{simon1976}%
  \BibitemOpen
  \bibfield  {author} {\bibinfo {author} {\bibfnamefont {B.}~\bibnamefont
  {Simon}},\ }\href@noop {} {\bibfield  {journal} {\bibinfo  {journal} {Ann.
  Phys.}\ }\textbf {\bibinfo {volume} {97}},\ \bibinfo {pages} {279} (\bibinfo
  {year} {1976})}\BibitemShut {NoStop}%
\bibitem [{\citenamefont {Volosniev}\ \emph {et~al.}(2014)\citenamefont
  {Volosniev}, \citenamefont {Fedorov}, \citenamefont {Jensen}, \citenamefont
  {Valiente},\ and\ \citenamefont {Zinner}}]{volosniev2014_nat}%
  \BibitemOpen
  \bibfield  {author} {\bibinfo {author} {\bibfnamefont {A.~G.}\ \bibnamefont
  {Volosniev}}, \bibinfo {author} {\bibfnamefont {D.~V.}\ \bibnamefont
  {Fedorov}}, \bibinfo {author} {\bibfnamefont {A.~S.}\ \bibnamefont {Jensen}},
  \bibinfo {author} {\bibfnamefont {M.}~\bibnamefont {Valiente}}, \ and\
  \bibinfo {author} {\bibfnamefont {N.~T.}\ \bibnamefont {Zinner}},\ }\href
  {\doibase 10.1038/ncomms6300} {\bibfield  {journal} {\bibinfo  {journal}
  {Nature Communications}\ }\textbf {\bibinfo {volume} {5}},\ \bibinfo {pages}
  {5300} (\bibinfo {year} {2014})}\BibitemShut {NoStop}%
\bibitem [{\citenamefont {Deuretzbacher}\ \emph {et~al.}(2014)\citenamefont
  {Deuretzbacher}, \citenamefont {Becker}, \citenamefont {Bjerlin},
  \citenamefont {Reimann},\ and\ \citenamefont {Santos}}]{santos2014}%
  \BibitemOpen
  \bibfield  {author} {\bibinfo {author} {\bibfnamefont {F.}~\bibnamefont
  {Deuretzbacher}}, \bibinfo {author} {\bibfnamefont {D.}~\bibnamefont
  {Becker}}, \bibinfo {author} {\bibfnamefont {J.}~\bibnamefont {Bjerlin}},
  \bibinfo {author} {\bibfnamefont {S.~M.}\ \bibnamefont {Reimann}}, \ and\
  \bibinfo {author} {\bibfnamefont {L.}~\bibnamefont {Santos}},\ }\href
  {\doibase 10.1103/PhysRevA.90.013611} {\bibfield  {journal} {\bibinfo
  {journal} {Phys. Rev. A}\ }\textbf {\bibinfo {volume} {90}},\ \bibinfo
  {pages} {013611} (\bibinfo {year} {2014})}\BibitemShut {NoStop}%
\bibitem [{\citenamefont {Volosniev}\ \emph {et~al.}(2015)\citenamefont
  {Volosniev}, \citenamefont {Petrosyan}, \citenamefont {Valiente},
  \citenamefont {Fedorov}, \citenamefont {Jensen},\ and\ \citenamefont
  {Zinner}}]{volosniev2015}%
  \BibitemOpen
  \bibfield  {author} {\bibinfo {author} {\bibfnamefont {A.~G.}\ \bibnamefont
  {Volosniev}}, \bibinfo {author} {\bibfnamefont {D.}~\bibnamefont
  {Petrosyan}}, \bibinfo {author} {\bibfnamefont {M.}~\bibnamefont {Valiente}},
  \bibinfo {author} {\bibfnamefont {D.~V.}\ \bibnamefont {Fedorov}}, \bibinfo
  {author} {\bibfnamefont {A.~S.}\ \bibnamefont {Jensen}}, \ and\ \bibinfo
  {author} {\bibfnamefont {N.~T.}\ \bibnamefont {Zinner}},\ }\href {\doibase
  10.1103/PhysRevA.91.023620} {\bibfield  {journal} {\bibinfo  {journal} {Phys.
  Rev. A}\ }\textbf {\bibinfo {volume} {91}},\ \bibinfo {pages} {023620}
  (\bibinfo {year} {2015})}\BibitemShut {NoStop}%
\bibitem [{\citenamefont {Massignan}\ \emph {et~al.}(2015)\citenamefont
  {Massignan}, \citenamefont {Levinsen},\ and\ \citenamefont
  {Parish}}]{massignan2015}%
  \BibitemOpen
  \bibfield  {author} {\bibinfo {author} {\bibfnamefont {P.}~\bibnamefont
  {Massignan}}, \bibinfo {author} {\bibfnamefont {J.}~\bibnamefont {Levinsen}},
  \ and\ \bibinfo {author} {\bibfnamefont {M.~M.}\ \bibnamefont {Parish}},\
  }\href {\doibase 10.1103/PhysRevLett.115.247202} {\bibfield  {journal}
  {\bibinfo  {journal} {Phys. Rev. Lett.}\ }\textbf {\bibinfo {volume} {115}},\
  \bibinfo {pages} {247202} (\bibinfo {year} {2015})}\BibitemShut {NoStop}%
\bibitem [{\citenamefont {Deuretzbacher}\ \emph {et~al.}(2017)\citenamefont
  {Deuretzbacher}, \citenamefont {Becker}, \citenamefont {Bjerlin},
  \citenamefont {Reimann},\ and\ \citenamefont {Santos}}]{santos2017}%
  \BibitemOpen
  \bibfield  {author} {\bibinfo {author} {\bibfnamefont {F.}~\bibnamefont
  {Deuretzbacher}}, \bibinfo {author} {\bibfnamefont {D.}~\bibnamefont
  {Becker}}, \bibinfo {author} {\bibfnamefont {J.}~\bibnamefont {Bjerlin}},
  \bibinfo {author} {\bibfnamefont {S.~M.}\ \bibnamefont {Reimann}}, \ and\
  \bibinfo {author} {\bibfnamefont {L.}~\bibnamefont {Santos}},\ }\href
  {\doibase 10.1103/PhysRevA.95.043630} {\bibfield  {journal} {\bibinfo
  {journal} {Phys. Rev. A}\ }\textbf {\bibinfo {volume} {95}},\ \bibinfo
  {pages} {043630} (\bibinfo {year} {2017})}\BibitemShut {NoStop}%
\bibitem [{\citenamefont {Volosniev}(2017)}]{volosniev2017}%
  \BibitemOpen
  \bibfield  {author} {\bibinfo {author} {\bibfnamefont {A.~G.}\ \bibnamefont
  {Volosniev}},\ }\href {\doibase 10.1007/s00601-017-1227-0} {\bibfield
  {journal} {\bibinfo  {journal} {Few-Body Systems}\ }\textbf {\bibinfo
  {volume} {58}},\ \bibinfo {pages} {54} (\bibinfo {year} {2017})}\BibitemShut
  {NoStop}%
\bibitem [{\citenamefont {Hodgson}\ and\ \citenamefont
  {Parkinson}(1984)}]{Hodgson_1984}%
  \BibitemOpen
  \bibfield  {author} {\bibinfo {author} {\bibfnamefont {R.~P.}\ \bibnamefont
  {Hodgson}}\ and\ \bibinfo {author} {\bibfnamefont {J.~B.}\ \bibnamefont
  {Parkinson}},\ }\href {\doibase 10.1088/0022-3719/17/18/014} {\bibfield
  {journal} {\bibinfo  {journal} {Journal of Physics C: Solid State Physics}\
  }\textbf {\bibinfo {volume} {17}},\ \bibinfo {pages} {3223} (\bibinfo {year}
  {1984})}\BibitemShut {NoStop}%
\bibitem [{\citenamefont {Recati}\ \emph {et~al.}(2005)\citenamefont {Recati},
  \citenamefont {Fuchs}, \citenamefont {Pe\ifmmode~\mbox{\c{c}}\else
  \c{c}\fi{}a},\ and\ \citenamefont {Zwerger}}]{zwerger2005}%
  \BibitemOpen
  \bibfield  {author} {\bibinfo {author} {\bibfnamefont {A.}~\bibnamefont
  {Recati}}, \bibinfo {author} {\bibfnamefont {J.~N.}\ \bibnamefont {Fuchs}},
  \bibinfo {author} {\bibfnamefont {C.~S.}\ \bibnamefont
  {Pe\ifmmode~\mbox{\c{c}}\else \c{c}\fi{}a}}, \ and\ \bibinfo {author}
  {\bibfnamefont {W.}~\bibnamefont {Zwerger}},\ }\href {\doibase
  10.1103/PhysRevA.72.023616} {\bibfield  {journal} {\bibinfo  {journal} {Phys.
  Rev. A}\ }\textbf {\bibinfo {volume} {72}},\ \bibinfo {pages} {023616}
  (\bibinfo {year} {2005})}\BibitemShut {NoStop}%
\bibitem [{\citenamefont {Fuchs}\ \emph {et~al.}(2007)\citenamefont {Fuchs},
  \citenamefont {Recati},\ and\ \citenamefont {Zwerger}}]{zwerger2007}%
  \BibitemOpen
  \bibfield  {author} {\bibinfo {author} {\bibfnamefont {J.~N.}\ \bibnamefont
  {Fuchs}}, \bibinfo {author} {\bibfnamefont {A.}~\bibnamefont {Recati}}, \
  and\ \bibinfo {author} {\bibfnamefont {W.}~\bibnamefont {Zwerger}},\ }\href
  {\doibase 10.1103/PhysRevA.75.043615} {\bibfield  {journal} {\bibinfo
  {journal} {Phys. Rev. A}\ }\textbf {\bibinfo {volume} {75}},\ \bibinfo
  {pages} {043615} (\bibinfo {year} {2007})}\BibitemShut {NoStop}%
\bibitem [{\citenamefont {Parisi}\ and\ \citenamefont
  {Giorgini}(2017)}]{parisi2017}%
  \BibitemOpen
  \bibfield  {author} {\bibinfo {author} {\bibfnamefont {L.}~\bibnamefont
  {Parisi}}\ and\ \bibinfo {author} {\bibfnamefont {S.}~\bibnamefont
  {Giorgini}},\ }\href {\doibase 10.1103/PhysRevA.95.023619} {\bibfield
  {journal} {\bibinfo  {journal} {Phys. Rev. A}\ }\textbf {\bibinfo {volume}
  {95}},\ \bibinfo {pages} {023619} (\bibinfo {year} {2017})}\BibitemShut
  {NoStop}%
\bibitem [{\citenamefont {Anderson}(1967)}]{Anderson1967}%
  \BibitemOpen
  \bibfield  {author} {\bibinfo {author} {\bibfnamefont {P.~W.}\ \bibnamefont
  {Anderson}},\ }\href {\doibase 10.1103/PhysRevLett.18.1049} {\bibfield
  {journal} {\bibinfo  {journal} {Phys. Rev. Lett.}\ }\textbf {\bibinfo
  {volume} {18}},\ \bibinfo {pages} {1049} (\bibinfo {year}
  {1967})}\BibitemShut {NoStop}%
\bibitem [{\citenamefont {Castella}\ and\ \citenamefont
  {Zotos}(1993)}]{zotos1993}%
  \BibitemOpen
  \bibfield  {author} {\bibinfo {author} {\bibfnamefont {H.}~\bibnamefont
  {Castella}}\ and\ \bibinfo {author} {\bibfnamefont {X.}~\bibnamefont
  {Zotos}},\ }\href {\doibase 10.1103/PhysRevB.47.16186} {\bibfield  {journal}
  {\bibinfo  {journal} {Phys. Rev. B}\ }\textbf {\bibinfo {volume} {47}},\
  \bibinfo {pages} {16186} (\bibinfo {year} {1993})}\BibitemShut {NoStop}%
\bibitem [{\citenamefont {Giraud}\ and\ \citenamefont
  {Combescot}(2009)}]{Combescot2009}%
  \BibitemOpen
  \bibfield  {author} {\bibinfo {author} {\bibfnamefont {S.}~\bibnamefont
  {Giraud}}\ and\ \bibinfo {author} {\bibfnamefont {R.}~\bibnamefont
  {Combescot}},\ }\href {\doibase 10.1103/PhysRevA.79.043615} {\bibfield
  {journal} {\bibinfo  {journal} {Phys. Rev. A}\ }\textbf {\bibinfo {volume}
  {79}},\ \bibinfo {pages} {043615} (\bibinfo {year} {2009})}\BibitemShut
  {NoStop}%
\bibitem [{\citenamefont {Mao}\ \emph {et~al.}(2016)\citenamefont {Mao},
  \citenamefont {Guan},\ and\ \citenamefont {Wu}}]{guan2016}%
  \BibitemOpen
  \bibfield  {author} {\bibinfo {author} {\bibfnamefont {R.}~\bibnamefont
  {Mao}}, \bibinfo {author} {\bibfnamefont {X.~W.}\ \bibnamefont {Guan}}, \
  and\ \bibinfo {author} {\bibfnamefont {B.}~\bibnamefont {Wu}},\ }\href
  {\doibase 10.1103/PhysRevA.94.043645} {\bibfield  {journal} {\bibinfo
  {journal} {Phys. Rev. A}\ }\textbf {\bibinfo {volume} {94}},\ \bibinfo
  {pages} {043645} (\bibinfo {year} {2016})}\BibitemShut {NoStop}%
\bibitem [{\citenamefont {Volosniev}\ and\ \citenamefont
  {Hammer}(2017)}]{Volosniev_2017a}%
  \BibitemOpen
  \bibfield  {author} {\bibinfo {author} {\bibfnamefont {A.~G.}\ \bibnamefont
  {Volosniev}}\ and\ \bibinfo {author} {\bibfnamefont {H.-W.}\ \bibnamefont
  {Hammer}},\ }\href {\doibase 10.1088/1367-2630/aa9011} {\bibfield  {journal}
  {\bibinfo  {journal} {New Journal of Physics}\ }\textbf {\bibinfo {volume}
  {19}},\ \bibinfo {pages} {113051} (\bibinfo {year} {2017})}\BibitemShut
  {NoStop}%
\bibitem [{\citenamefont {Vansant}\ \emph {et~al.}(1994)\citenamefont
  {Vansant}, \citenamefont {Smondyrev}, \citenamefont {Peeters},\ and\
  \citenamefont {Devreese}}]{Vansant_1994}%
  \BibitemOpen
  \bibfield  {author} {\bibinfo {author} {\bibfnamefont {P.}~\bibnamefont
  {Vansant}}, \bibinfo {author} {\bibfnamefont {M.~A.}\ \bibnamefont
  {Smondyrev}}, \bibinfo {author} {\bibfnamefont {F.~M.}\ \bibnamefont
  {Peeters}}, \ and\ \bibinfo {author} {\bibfnamefont {J.~T.}\ \bibnamefont
  {Devreese}},\ }\href {\doibase 10.1088/0305-4470/27/23/035} {\bibfield
  {journal} {\bibinfo  {journal} {Journal of Physics A: Mathematical and
  General}\ }\textbf {\bibinfo {volume} {27}},\ \bibinfo {pages} {7925}
  (\bibinfo {year} {1994})}\BibitemShut {NoStop}%
\bibitem [{\citenamefont {Sowi{\'n}ski}\ and\ \citenamefont
  {Garc{\'i}a-March}()}]{sowinski2019}%
  \BibitemOpen
  \bibfield  {author} {\bibinfo {author} {\bibfnamefont {T.}~\bibnamefont
  {Sowi{\'n}ski}}\ and\ \bibinfo {author} {\bibfnamefont {M.~{\'A}.}\
  \bibnamefont {Garc{\'i}a-March}},\ }\href@noop {} {\bibinfo  {journal} {{\bf
  arXiv}:1903.12189}\ }\BibitemShut {NoStop}%
\bibitem [{\citenamefont {Flicker}\ and\ \citenamefont
  {Lieb}(1967)}]{lieb1967a}%
  \BibitemOpen
\bibfield  {journal} {  }\bibfield  {author} {\bibinfo {author} {\bibfnamefont
  {M.}~\bibnamefont {Flicker}}\ and\ \bibinfo {author} {\bibfnamefont {E.~H.}\
  \bibnamefont {Lieb}},\ }\href {\doibase 10.1103/PhysRev.161.179} {\bibfield
  {journal} {\bibinfo  {journal} {Phys. Rev.}\ }\textbf {\bibinfo {volume}
  {161}},\ \bibinfo {pages} {179} (\bibinfo {year} {1967})}\BibitemShut
  {NoStop}%
\bibitem [{\citenamefont {Rammelm\"uller}\ \emph
  {et~al.}(2017{\natexlab{b}})\citenamefont {Rammelm\"uller}, \citenamefont
  {Porter}, \citenamefont {Drut},\ and\ \citenamefont
  {Braun}}]{Rammelmuller2017}%
  \BibitemOpen
  \bibfield  {author} {\bibinfo {author} {\bibfnamefont {L.}~\bibnamefont
  {Rammelm\"uller}}, \bibinfo {author} {\bibfnamefont {W.~J.}\ \bibnamefont
  {Porter}}, \bibinfo {author} {\bibfnamefont {J.~E.}\ \bibnamefont {Drut}}, \
  and\ \bibinfo {author} {\bibfnamefont {J.}~\bibnamefont {Braun}},\ }\href
  {\doibase 10.1103/PhysRevD.96.094506} {\bibfield  {journal} {\bibinfo
  {journal} {Phys. Rev. D}\ }\textbf {\bibinfo {volume} {96}},\ \bibinfo
  {pages} {094506} (\bibinfo {year} {2017}{\natexlab{b}})}\BibitemShut
  {NoStop}%
\bibitem [{\citenamefont {Rammelmüller}\ \emph {et~al.}(2018)\citenamefont
  {Rammelmüller}, \citenamefont {Drut},\ and\ \citenamefont
  {Braun}}]{Rammelm_ller_2018}%
  \BibitemOpen
  \bibfield  {author} {\bibinfo {author} {\bibfnamefont {L.}~\bibnamefont
  {Rammelmüller}}, \bibinfo {author} {\bibfnamefont {J.~E.}\ \bibnamefont
  {Drut}}, \ and\ \bibinfo {author} {\bibfnamefont {J.}~\bibnamefont {Braun}},\
  }\href {\doibase 10.1088/1742-6596/1041/1/012006} {\bibfield  {journal}
  {\bibinfo  {journal} {Journal of Physics: Conference Series}\ }\textbf
  {\bibinfo {volume} {1041}},\ \bibinfo {pages} {012006} (\bibinfo {year}
  {2018})}\BibitemShut {NoStop}%
\bibitem [{\citenamefont {Girardeau}(1960)}]{Girardeau1960}%
  \BibitemOpen
  \bibfield  {author} {\bibinfo {author} {\bibfnamefont {M.}~\bibnamefont
  {Girardeau}},\ }\href {\doibase 10.1063/1.1703687} {\bibfield  {journal}
  {\bibinfo  {journal} {Journal of Mathematical Physics}\ }\textbf {\bibinfo
  {volume} {1}},\ \bibinfo {pages} {516} (\bibinfo {year} {1960})},\ \Eprint
  {http://arxiv.org/abs/https://doi.org/10.1063/1.1703687}
  {https://doi.org/10.1063/1.1703687} \BibitemShut {NoStop}%
\bibitem [{\citenamefont {Grusdt}\ \emph {et~al.}(2017)\citenamefont {Grusdt},
  \citenamefont {Astrakharchik},\ and\ \citenamefont {Demler}}]{Grusdt_2017}%
  \BibitemOpen
  \bibfield  {author} {\bibinfo {author} {\bibfnamefont {F.}~\bibnamefont
  {Grusdt}}, \bibinfo {author} {\bibfnamefont {G.~E.}\ \bibnamefont
  {Astrakharchik}}, \ and\ \bibinfo {author} {\bibfnamefont {E.}~\bibnamefont
  {Demler}},\ }\href {\doibase 10.1088/1367-2630/aa8a2e} {\bibfield  {journal}
  {\bibinfo  {journal} {New Journal of Physics}\ }\textbf {\bibinfo {volume}
  {19}},\ \bibinfo {pages} {103035} (\bibinfo {year} {2017})}\BibitemShut
  {NoStop}%
\bibitem [{\citenamefont {Pastukhov}(2017)}]{Pastukhov2017}%
  \BibitemOpen
  \bibfield  {author} {\bibinfo {author} {\bibfnamefont {V.}~\bibnamefont
  {Pastukhov}},\ }\href {\doibase 10.1103/PhysRevA.96.043625} {\bibfield
  {journal} {\bibinfo  {journal} {Phys. Rev. A}\ }\textbf {\bibinfo {volume}
  {96}},\ \bibinfo {pages} {043625} (\bibinfo {year} {2017})}\BibitemShut
  {NoStop}%
\bibitem [{\citenamefont {Kain}\ and\ \citenamefont {Ling}(2018)}]{kain2018}%
  \BibitemOpen
  \bibfield  {author} {\bibinfo {author} {\bibfnamefont {B.}~\bibnamefont
  {Kain}}\ and\ \bibinfo {author} {\bibfnamefont {H.~Y.}\ \bibnamefont
  {Ling}},\ }\href {\doibase 10.1103/PhysRevA.98.033610} {\bibfield  {journal}
  {\bibinfo  {journal} {Phys. Rev. A}\ }\textbf {\bibinfo {volume} {98}},\
  \bibinfo {pages} {033610} (\bibinfo {year} {2018})}\BibitemShut {NoStop}%
\bibitem [{\citenamefont {Mistakidis}\ \emph
  {et~al.}(2019{\natexlab{c}})\citenamefont {Mistakidis}, \citenamefont
  {Katsimiga}, \citenamefont {Koutentakis}, \citenamefont {Busch},\ and\
  \citenamefont {Schmelcher}}]{mistakidis2019_quench}%
  \BibitemOpen
  \bibfield  {author} {\bibinfo {author} {\bibfnamefont {S.~I.}\ \bibnamefont
  {Mistakidis}}, \bibinfo {author} {\bibfnamefont {G.~C.}\ \bibnamefont
  {Katsimiga}}, \bibinfo {author} {\bibfnamefont {G.~M.}\ \bibnamefont
  {Koutentakis}}, \bibinfo {author} {\bibfnamefont {T.}~\bibnamefont {Busch}},
  \ and\ \bibinfo {author} {\bibfnamefont {P.}~\bibnamefont {Schmelcher}},\
  }\href {\doibase 10.1103/PhysRevLett.122.183001} {\bibfield  {journal}
  {\bibinfo  {journal} {Phys. Rev. Lett.}\ }\textbf {\bibinfo {volume} {122}},\
  \bibinfo {pages} {183001} (\bibinfo {year} {2019}{\natexlab{c}})}\BibitemShut
  {NoStop}%
\bibitem [{\citenamefont {Paeckel}\ \emph {et~al.}(2019)\citenamefont
  {Paeckel}, \citenamefont {Köhler}, \citenamefont {Swoboda}, \citenamefont
  {Manmana}, \citenamefont {Schollwöck},\ and\ \citenamefont
  {Hubig}}]{Paeckel:2019yjf}%
  \BibitemOpen
  \bibfield  {author} {\bibinfo {author} {\bibfnamefont {S.}~\bibnamefont
  {Paeckel}}, \bibinfo {author} {\bibfnamefont {T.}~\bibnamefont {Köhler}},
  \bibinfo {author} {\bibfnamefont {A.}~\bibnamefont {Swoboda}}, \bibinfo
  {author} {\bibfnamefont {S.~R.}\ \bibnamefont {Manmana}}, \bibinfo {author}
  {\bibfnamefont {U.}~\bibnamefont {Schollwöck}}, \ and\ \bibinfo {author}
  {\bibfnamefont {C.}~\bibnamefont {Hubig}},\ }\href@noop {} {\  (\bibinfo
  {year} {2019})},\ \Eprint {http://arxiv.org/abs/1901.05824}
  {arXiv:1901.05824} \BibitemShut {NoStop}%
\bibitem [{\citenamefont {Schollw\"ock}(2005)}]{RevModPhys.77.259}%
  \BibitemOpen
  \bibfield  {author} {\bibinfo {author} {\bibfnamefont {U.}~\bibnamefont
  {Schollw\"ock}},\ }\href {\doibase 10.1103/RevModPhys.77.259} {\bibfield
  {journal} {\bibinfo  {journal} {Rev. Mod. Phys.}\ }\textbf {\bibinfo {volume}
  {77}},\ \bibinfo {pages} {259} (\bibinfo {year} {2005})}\BibitemShut
  {NoStop}%
\bibitem [{\citenamefont {Gull}\ \emph {et~al.}(2011)\citenamefont {Gull},
  \citenamefont {Millis}, \citenamefont {Lichtenstein}, \citenamefont
  {Rubtsov}, \citenamefont {Troyer},\ and\ \citenamefont
  {Werner}}]{RevModPhys.83.349}%
  \BibitemOpen
  \bibfield  {author} {\bibinfo {author} {\bibfnamefont {E.}~\bibnamefont
  {Gull}}, \bibinfo {author} {\bibfnamefont {A.~J.}\ \bibnamefont {Millis}},
  \bibinfo {author} {\bibfnamefont {A.~I.}\ \bibnamefont {Lichtenstein}},
  \bibinfo {author} {\bibfnamefont {A.~N.}\ \bibnamefont {Rubtsov}}, \bibinfo
  {author} {\bibfnamefont {M.}~\bibnamefont {Troyer}}, \ and\ \bibinfo {author}
  {\bibfnamefont {P.}~\bibnamefont {Werner}},\ }\href {\doibase
  10.1103/RevModPhys.83.349} {\bibfield  {journal} {\bibinfo  {journal} {Rev.
  Mod. Phys.}\ }\textbf {\bibinfo {volume} {83}},\ \bibinfo {pages} {349}
  (\bibinfo {year} {2011})}\BibitemShut {NoStop}%
\bibitem [{\citenamefont {Pasek}\ and\ \citenamefont {Orso}(2019)}]{orso2019}%
  \BibitemOpen
  \bibfield  {author} {\bibinfo {author} {\bibfnamefont {M.}~\bibnamefont
  {Pasek}}\ and\ \bibinfo {author} {\bibfnamefont {G.}~\bibnamefont {Orso}},\
  }\href@noop {} {\  (\bibinfo {year} {2019})},\ \Eprint
  {http://arxiv.org/abs/1910.03569} {arXiv:1910.03569} \BibitemShut {NoStop}%
\bibitem [{\citenamefont {Oelkers}\ \emph {et~al.}(2006)\citenamefont
  {Oelkers}, \citenamefont {Batchelor}, \citenamefont {Bortz},\ and\
  \citenamefont {Guan}}]{oelkers2006}%
  \BibitemOpen
  \bibfield  {author} {\bibinfo {author} {\bibfnamefont {N.}~\bibnamefont
  {Oelkers}}, \bibinfo {author} {\bibfnamefont {M.~T.}\ \bibnamefont
  {Batchelor}}, \bibinfo {author} {\bibfnamefont {M.}~\bibnamefont {Bortz}}, \
  and\ \bibinfo {author} {\bibfnamefont {X.-W.}\ \bibnamefont {Guan}},\ }\href
  {\doibase 10.1088/0305-4470/39/5/005} {\bibfield  {journal} {\bibinfo
  {journal} {Journal of Physics A: Mathematical and General}\ }\textbf
  {\bibinfo {volume} {39}},\ \bibinfo {pages} {1073} (\bibinfo {year}
  {2006})}\BibitemShut {NoStop}%
\end{thebibliography}%

\end{document}